\PassOptionsToPackage{table}{xcolor}
\documentclass[sigconf,nonacm]{acmart}
\setcopyright{none}
\usepackage{booktabs}
\usepackage{multirow}
\usepackage{siunitx}
\usepackage[autostyle]{csquotes}
\usepackage{tikz}
\usetikzlibrary{positioning,calc,arrows.meta}
\usepackage{pgfplots}
\usepackage{pgfplotstable}
\pgfplotsset{compat=1.18}
\usepackage{pgfplotstable}
\usepackage{standalone}
\usepackage{adjustbox}
\usepackage{xcolor}
\usepackage{pgfplots}
\pgfplotsset{compat=1.18}
\usepgfplotslibrary{groupplots,fillbetween}
\usetikzlibrary{patterns,arrows.meta,positioning,calc,fit,matrix,backgrounds}

\pgfplotsset{table/col sep=comma}
\IfFileExists{figures/figstyle.tex}{}{}

\definecolor{arGray}{HTML}{555555}
\definecolor{mdlmOrange}{HTML}{D55E00}
\definecolor{bdeightBlue}{HTML}{0072B2}
\definecolor{bdfourGreen}{HTML}{009E73}
\definecolor{okYellow}{HTML}{F0E442}
\definecolor{okSky}{HTML}{56B4E9}
\definecolor{okPurple}{HTML}{CC79A7}
\colorlet{figInk}{black!87}
\colorlet{figMuted}{black!46}
\colorlet{figBand}{black!10}
\colorlet{figGrid}{black!8}

\pgfplotsset{
  figaxis/.style={
    scale only axis,
    axis x line*=bottom, axis y line*=left,
    axis line style={draw=figMuted, line width=0.4pt},
    tick style={draw=figMuted, line width=0.4pt},
    major tick length=2pt,
    ymajorgrids, grid style={draw=figGrid, line width=0.4pt},
    tick label style={font=\tiny, text=figInk},
    label style={font=\scriptsize, text=figInk},
    xlabel style={align=center},
    title style={font=\scriptsize, text=figInk, at={(0,1)}, anchor=south west,
      yshift=-1pt},
    legend style={font=\tiny, draw=none, fill=none, text=figInk},
    legend cell align=left,
    yticklabel style={/pgf/number format/fixed},
  },
  paradigmx/.style={
    xmin=-0.6, xmax=3.6, xtick={0,1,2,3},
    xticklabels={AR,MDLM,BD8,BD4}, xmajorgrids=false,
  },
  solidbar/.style={ybar, bar width=0.34, bar shift=-0.19, draw=none,
    fill=#1},
  hatchbar/.style={ybar, bar width=0.34, bar shift=0.19, draw=#1,
    line width=0.5pt, pattern=north east lines, pattern color=#1},
  dotbar/.style={ybar, bar width=0.34, bar shift=0.19, draw=#1,
    line width=0.5pt, pattern=dots, pattern color=#1},
  pmark arGray/.style={mark=*},
  pmark mdlmOrange/.style={mark=square*},
  pmark bdeightBlue/.style={mark=triangle*},
  pmark bdfourGreen/.style={mark=diamond*},
  paradigmline/.style={color=#1, line width=0.9pt, pmark #1,
    mark size=1.3pt, mark options={solid, fill=#1}},
  percent y/.style={scaled y ticks=false, yticklabel={\pgfmathparse{100*\tick}%
    \pgfmathprintnumber[fixed, precision=0]{\pgfmathresult}}},
  percent x/.style={scaled x ticks=false, xticklabel={\pgfmathparse{100*\tick}%
    \pgfmathprintnumber[fixed, precision=0]{\pgfmathresult}}},
}

\hypersetup{colorlinks, linkcolor=blue, citecolor=blue, urlcolor=blue}
\usepackage[capitalize,nameinlink]{cleveref}

\makeatletter
\let\colfigure\figure \let\endcolfigure\endfigure
\renewenvironment{figure}[1][]{\@dblfloat{figure}[t]}{\end@dblfloat}
\makeatother

\newcommand{\codeurl}{\footnote{Released upon acceptance.}}
\newcommand{\dataurl}{\footnote{Released upon acceptance.}}

\definecolor{arGray}{HTML}{555555}
\definecolor{bdfourGreen}{HTML}{009E73}
\definecolor{bdeightBlue}{HTML}{0072B2}
\definecolor{mdlmOrange}{HTML}{D55E00}

\newcommand{\AR}{AR}
\newcommand{\BDfour}{BD4}
\newcommand{\BDeight}{BD8}
\newcommand{\MDLM}{MDLM}
\newcommand{\RQ}{RQ}
\newcommand{\PQ}{PQ}
\newcommand{\NQ}{NQ320K}
\newcommand{\MSM}{MS300K}

\newcommand{\tablefont}{\footnotesize\setlength{\tabcolsep}{1.5pt}}
\newcommand{\tablesep}{\setlength{\tabcolsep}{4pt}}
\newcommand{\armscolw}{0.6cm}
\newif\ifcolortables
\colortablesfalse
\newcommand{\pshape}[1]{%
  \ifcsname pshape@#1\endcsname\csname pshape@#1\endcsname
  \else\fill (0,0) circle (0.35ex);\fi}
\expandafter\def\csname pshape@arGray\endcsname{\fill (0,0) circle (0.35ex);}
\expandafter\def\csname pshape@mdlmOrange\endcsname{\fill (-0.32ex,-0.32ex) rectangle (0.32ex,0.32ex);}
\expandafter\def\csname pshape@bdeightBlue\endcsname{\fill (90:0.45ex) -- (210:0.45ex) -- (330:0.45ex) -- cycle;}
\expandafter\def\csname pshape@bdfourGreen\endcsname{\fill (0,0.45ex) -- (0.4ex,0) -- (0,-0.45ex) -- (-0.4ex,0) -- cycle;}
\newcommand{\pdot}[1]{\ifcolortables
  \tikz[baseline=-0.6ex, #1]{\pshape{#1}}\,%
  \else\tikz[baseline=-0.6ex]{\pshape{#1}}\,\fi}

\newcommand{\ci}[2]{\textcolor{gray}{\scriptsize[#1,\,#2]}}
\newcommand{\best}[1]{\bfseries #1}
\newcommand{\heat}[1]{\cellcolor{black!\fpeval{round(0.35*#1)}}#1}

\begin{document}

\title{Disentangling Paradigm, Identifier, and Decoding in Generative Retrieval}

\author{Hicham Randrianarivo}
\orcid{0009-0006-1764-5253}
\affiliation{\institution{Artefact Research Center}\country{France}}
\email{hicham.randrianarivo@artefact.com}

\author{Logan Renaud}
\affiliation{\institution{Artefact Research Center}\country{France}}
\email{logan.renaud@artefact.com}

\author{Alexia Allal}
\affiliation{\institution{Artefact Research Center}\country{France}}
\email{alexia.allal@artefact.com}

\renewcommand{\shortauthors}{Randrianarivo et al.}

\begin{abstract}
Generative retrieval trains a language model to generate the identifier
of a relevant document. Recent work replaces the autoregressive decoder
with diffusion, but changes identifiers, training recipe and decoding
at once, so differences cannot be credited to the paradigm. On NQ320K and MS300K, we train autoregressive,
masked-diffusion and block-diffusion models with residual-quantised,
product-quantised and random identifiers. With identifier length and
training budget fixed, we decode each model in several ways.
Decoding alone moves a diffusion model's Hit@1 by 6.6 to 13.7 points. Our reference diffusion decoding, generate-and-match, generates an identifier, then retrieves the closest corpus
identifiers. The generated identifier is right for 14--21\% of NQ320K queries. We test one-pass scoring to decode diffusion retrievers: the model
reads a fully masked identifier once, and each document is scored by its
codes' probabilities. It matches or beats generate-and-match in 11 of 12 settings.
Autoregressive models still lead in Hit@1; on NQ320K, the lead comes from the model, not beam search. Starting from one
sampled identifier, one-pass scoring removes 46--83\% of masked diffusion's
deficit to beam search; from generate-and-match, at most a quarter. On NQ320K, every paradigm largely memorises which
identifier answers which query: random identifiers keep 83--90\% of the Hit@1 of residual-quantised ones. There, product-quantised identifiers lead
residual-quantised ones by 3.4 points in the autoregressive model and by
$-0.7$ to $+3.6$ in diffusion models; across decodings, AR's gap exceeds diffusion's by 1.5--2.3 points, around our $\pm2$-point threshold. Paradigm comparisons must report each paradigm at its own recipe and best decoding.

\end{abstract}

\ccsdesc[500]{Information systems~Retrieval models and ranking}

\keywords{Generative retrieval, Document identifiers, Diffusion language
  models, Decoding strategies, Semantic IDs, Controlled comparison}

\maketitle

\section{Introduction}
\label{sec:intro}

\begin{figure}[t]
  \centering
  \begin{tikzpicture}[x=1cm, y=1cm, font=\scriptsize]
  \tikzset{
    ptitle/.style={font=\small\bfseries, text=figInk, anchor=north},
    ovbox/.style={draw=figInk, line width=0.5pt, rounded corners=1.5pt,
      fill=white, inner sep=2pt, text=figInk, font=\tiny, align=center,
      minimum height=0.42cm},
    pbox/.style={ovbox, line width=1pt, font=\tiny\bfseries, inner sep=1.5pt},
    family/.style={draw=figInk, fill=white, line width=1pt, rounded corners=2pt,
      inner sep=2.5pt},
    flabel/.style={font=\tiny\bfseries, text=figInk, inner sep=0.5pt},
    finding/.style={font=\tiny\itshape, text=figInk, align=center,
      anchor=north},
    ovarrow/.style={-{Latex[length=3pt]}, draw=figInk, line width=0.6pt}
  }
  \begin{scope}[on background layer]
    \fill[okYellow!30, rounded corners=3pt] (0,0) rectangle (2.7,3.6);
    \fill[okSky!18, rounded corners=3pt] (3.2,0) rectangle (6.3,3.6);
    \fill[figBand, rounded corners=3pt] (6.8,0) rectangle (10.6,3.6);
  \end{scope}
  \node[ptitle] at (1.35,3.52) {Identifier};
  \node[ptitle] at (4.75,3.52) {Paradigm};
  \node[ptitle] at (8.7,3.52) {Decoding};

  \node[ovbox, text width=2.1cm] (ids) at (1.35,2.5) {RQ $\cdot$ PQ};
  \node[ovbox, text width=2.1cm] (ctl) at (1.35,1.35)
    {controls:\\ permuted RQ $\cdot$ random};

  \matrix (arm) [column sep=0.14cm, inner sep=0pt, ampersand replacement=\&]
    at (4.75,2.5) {
    \node[pbox, draw=arGray, fill=arGray!15] {T5}; \\
  };
  \begin{scope}[on background layer]
    \node[family, fit=(arm), label={[flabel]above:Autoregressive}] (arf) {};
  \end{scope}
  \matrix (dm) [column sep=0.14cm, inner sep=0pt, ampersand replacement=\&]
    at (4.75,1.35) {
    \node[pbox, draw=mdlmOrange, fill=mdlmOrange!15] {MDLM}; \&
    \node[pbox, draw=bdeightBlue, fill=bdeightBlue!15] {BD8}; \&
    \node[pbox, draw=bdfourGreen, fill=bdfourGreen!15] {BD4}; \\
  };
  \begin{scope}[on background layer]
    \node[family, fit=(dm), label={[flabel]above:Diffusion}] (dif) {};
  \end{scope}

  \node[ovbox, line width=1pt] (beam) at (8.7,2.5) {trie beam};
  \matrix (dd) [column sep=0.1cm, inner sep=0pt, ampersand replacement=\&]
    at (8.7,1.35) {
    \node[ovbox, line width=1pt] {generate +\\code matching}; \&
    \node[font=\tiny, text=figMuted] {or}; \&
    \node[ovbox, line width=1pt] {one-pass\\scoring}; \\
  };
  \begin{scope}[on background layer]
    \node[family, fit=(dd)] (decf) {};
  \end{scope}

  \node[ovbox, minimum width=0.9cm, minimum height=1.75cm] (out)
    at (11.4,1.925) {docid};

  \draw[ovarrow] (2.7,1.925) -- node[above, font=\tiny, text=figInk,
    inner sep=1pt] {$\times$} (3.2,1.925);
  \draw[ovarrow] (arf.east) -- (beam.west);
  \draw[ovarrow] (dif.east) -- (decf.west);
  \draw[ovarrow] (beam.east) -- (beam.east -| out.west);
  \draw[ovarrow] (decf.east) -- (decf.east -| out.west);

  \node[finding] at (1.35,0.75) {vary the docid,\\same paradigm};
  \node[finding] at (4.75,0.75) {vary the paradigm,\\same docid};
  \node[finding] at (8.7,0.75) {vary the decoding,\\same trained model};

  \node[text=figMuted, font=\tiny, anchor=north] at (5.3,-0.1)
    {held fixed: 16$\times$512 codes $\cdot$ NQ320K, MS300K $\cdot$ 16M examples};
\end{tikzpicture}
  \caption{Each factor is varied while the other two stay fixed. Every
    paradigm is trained on every docid; each trained model is then decoded
    in several ways. Italics: the comparison that isolates each factor.}
  \label{fig:overview}
\end{figure}
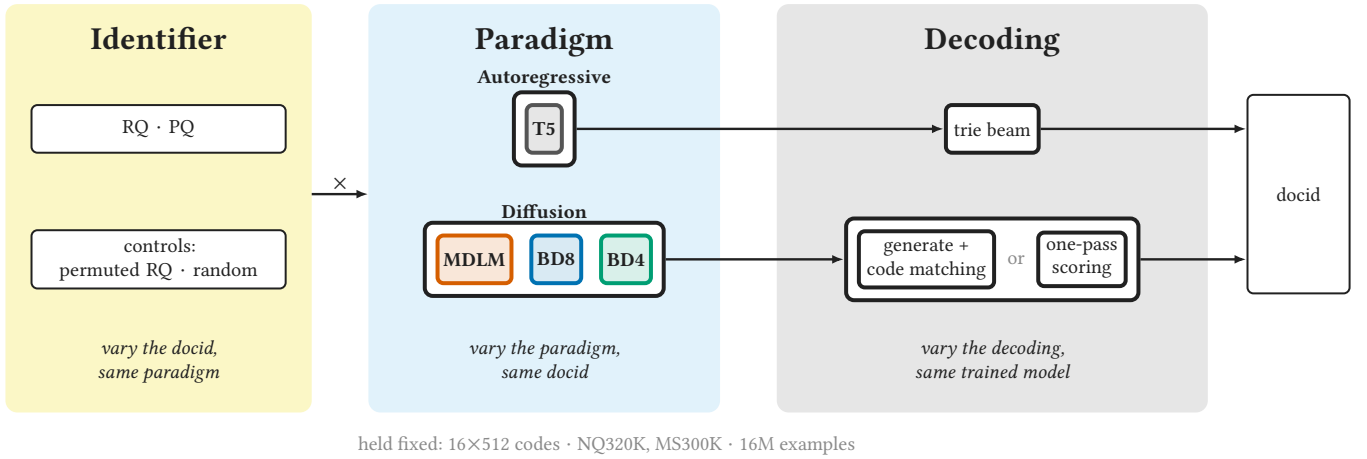

Generative retrieval (GR) maps a query directly to the identifier (docid)
of a relevant document. A docid is a short sequence of codes, and a
language model learns to generate it. Most GR models generate the docid
with an autoregressive (AR) decoder, one code at a time from left to right.
Recent work uses diffusion instead, in document
retrieval~\cite{zhao2025diffugr} and in
recommendation~\cite{liu2025diffgrm,shi2025lladarec,shah2025maskgr}. A
diffusion decoder starts from a fully masked docid and fills in its codes
over several steps. These works compare diffusion with AR, but they change
the identifier, the training settings and the decoding at the same time. A
difference in Hit@1 therefore cannot be credited to the generation
paradigm (AR or diffusion) alone.

We separate these three factors (\cref{fig:overview}). On NQ320K and
MS300K, we train four paradigms: AR, masked diffusion (MDLM), and block
diffusion with blocks of 4 or 8 codes (BD4, BD8). Each paradigm learns
several kinds of docids, from semantic to random. The semantic ones, RQ
and PQ, quantise a document embedding by residual or product
quantisation. Docid length, vocabulary and training budget are
fixed. We ask three questions.
\textbf{RQ1}: what does each paradigm learn? \textbf{RQ2}: what does
diffusion generate, and which decoding finds the right document?
\textbf{RQ3}: does the identifier favour a paradigm?

\Cref{fig:overview} shows how the study isolates each factor. We
train every paradigm on every docid, then keep these models fixed and
decode each one in several ways, from one sampled docid to scoring every
corpus docid in one forward pass.

This paper makes three contributions. First, a controlled design crosses
paradigm, identifier and decoding on the same trained models. With it, we
show that decoding alone moves a diffusion model's Hit@1 by 6.6 to 13.7 points.
Second, we turn one-pass scoring, a rule used by parallel-decoding
recommenders~\cite{hou2025rpg}, into a decoding for diffusion retrievers: one forward pass on a fully masked docid scores every corpus
docid. It matches or beats generate-and-match, which generates a docid
and reranks the closest corpus docids, in 11 of 12 settings. Third, we list what
a paradigm comparison should report. We release the code, splits and
docids.

\section{Related Work}
\label{sec:related}

\textbf{Generative retrieval.} DSI~\cite{tay2022dsi} casts
retrieval as sequence-to-sequence generation. A T5 model maps a query to
the identifier of a relevant document, and beam search in a trie of
valid docids returns a ranking. Docids range from text, such as the
n-grams of SEAL~\cite{bevilacqua2022seal}, to semantic codes built from
document embeddings: hierarchical k-means in DSI and
NCI~\cite{wang2022nci}, product quantisation in
DDRO~\cite{mekonnen2025ddro}, and residual quantisation in
TIGER~\cite{rajput2023tiger,singh2023semanticids}. Later work improves the codes, the training data and the
objective~\cite{sun2023genret,lee2023glen,zeng2023ripor,zhuang2022dsiqg,li2024ltrgr,phatak2025diversity,askari2026fewshot}. All decode autoregressively. Studies
of docid construction show that the code changes the results with the
model held fixed~\cite{valeau2026hierarchical,chen2026goodsid}. They keep
the paradigm fixed; we cross the identifier with the paradigm.

\textbf{Diffusion language models.} Discrete
diffusion~\cite{austin2021d3pm,lou2024sedd} corrupts a token sequence and
learns to undo the corruption. Masked diffusion
(MDLM~\cite{sahoo2024mdlm}) trains with a reweighted masked
language-modelling loss. It decodes by unmasking positions in any order.
Block diffusion (BD3-LM~\cite{arriola2025bd3lm}) decodes blocks from left
to right and the positions of a block in parallel. Comparisons with AR depend on
decoding. A reported diffusion advantage in language modelling came from
inaccurate sampling~\cite{zheng2025masked}. Diffusion models win on
per-token error but lose on whole-sequence error~\cite{feng2025theoretical}.
We therefore vary decoding on fixed checkpoints.

\textbf{Diffusion for generative retrieval.} DiffuGR~\cite{zhao2025diffugr} retrieves
documents with diffusion models built on LLaDA~\cite{nie2025llada} and
Dream~\cite{ye2025dream}, over RQ and text docids. It compares them with AR systems, each with its own docids,
training data and decoding. MaskGR~\cite{shah2025maskgr} compares masked
diffusion with an AR recommender. It tests random codes for diffusion
only. Diffusion recommenders~\cite{liu2025diffgrm,shi2025lladarec,mu2026masked}
argue, with ablations, that independent, PQ-like codes suit parallel
decoders. EPIC~\cite{tran2026epic} scores complete items during
denoising, because committing tokens one at a time can drop the target.
In all of them, the code, the recipe and the decoder change together, so
a gain cannot be credited to the paradigm. We cross these factors and test
whether PQ codes favour parallel decoders.

\textbf{Decoding.} PAG (Planning Ahead in Generative retrieval) scores
every document in one forward pass by summing the weights of a
second, lexical docid, and uses these scores as priors for AR beam
search~\cite{zeng2024pag}. Wu et al.\ show that AR GR scores a document
like multi-vector dense retrieval, with query vectors that depend on the
docid prefix~\cite{wu2024gr_mvdr}. RPG scores every item by summing
per-digit log-probabilities from one pass of a parallel-prediction
model~\cite{hou2025rpg}. One-pass scoring reads the query with
every docid level masked, and scores each docid by the probabilities of
its codes. We apply this rule to the denoiser of a trained diffusion model, with no
extra head, beam or denoising.

\textbf{Controlled analyses.} The closest analyses study AR only: memorisation,
new documents and robustness~\cite{chen2023understanding,kim2024dynamic,zhang2025replication,liu2025robustness,mekonnen2026lost}. On NQ320K, naive
and semantic docids reach the same Hit@1 for
AR~\cite{pradeep2023scale,nicolas2026redsi}. Scaling studies vary model,
corpus and inference budget for AR~\cite{pradeep2023scale,cai2025scaling}.
Controlled studies in neural IR vary one factor at a
time~\cite{lassance2023pretraining,lupart2023msshift,xion2026bias}. None
crosses the generation paradigm with the identifier. We do so at one model
scale, on the same corpora and budget.

\section{Design Space}
\label{sec:method}

A generative retriever has three parts: the docid it generates, the
model that generates it (its paradigm, how it reads the query, how it is
trained), and the decoding that turns the model's output into a ranked
list. We describe the options we test for each part.

\label{sec:meth-ids}
\textbf{Identifiers.} A docid gives each document $d$ a code
$c(d) = (c_1(d),\dots,c_L(d)) \in [K]^L$: $L$ levels, each taking one of
$K$ codes. Each (level, code) pair is a new token, so the docids expand the model's
vocabulary by $L{\times}K$ tokens~\cite{rajput2023tiger,mekonnen2025ddro}. All our docids share
this shape and vocabulary; they differ only in what a level means. Two are
built from a document embedding $e(d) \in \mathbb{R}^D$, with one k-means
codebook $\{\mu_{l,k}\}_{k=1}^{K}$ per level. Two are controls.

\textbf{RQ} (residual quantisation) encodes what the earlier levels left
over. It starts from $r_0(d) = e(d)$ and fits level $l$'s codebook on the
corpus residuals $r_{l-1}$: $c_l(d) = \arg\min_k \lVert r_{l-1}(d) -
\mu_{l,k} \rVert$ and $r_l(d) = r_{l-1}(d) - \mu_{l,c_l(d)}$. RQ levels go from coarse to fine. A level only makes sense given the
earlier ones, and similar documents share prefixes.

\textbf{PQ} (product quantisation) cuts $e(d) = [e^1(d); \dots; e^L(d)]$
into $L$ slices of $D/L$ dimensions and fits level $l$'s codebook on
slice $l$ of every document: $c_l(d) = \arg\min_k \lVert e^l(d) -
\mu_{l,k} \rVert$. PQ levels are independent. Each makes sense alone, and none refines
another. Similar documents share many levels, at any position, and some
share a whole code.

\textbf{Random} codes draw each $c_l(d)$ uniformly from $[K]$, once per
document; no two documents share a code. They have the shape of RQ and
PQ but no semantic meaning, so a model can only memorise which queries
map to them.

\textbf{Permuted} RQ codes reorder the levels of each RQ code:
$c^{\pi}_l(d) = c_{\pi(l)}(d)$, with $\pi$ the reversal or one fixed
random permutation. The codes and collisions stay the same; only the
coarse-to-fine order is lost.

\label{sec:meth-models}
\textbf{Paradigms.} The paradigms differ in the order in which they
commit the $L$ levels. AR commits one level at a time, from left to right.
Block diffusion commits blocks of $k$ levels from left to right, and the
levels inside a block in parallel. Masked diffusion (MDLM) treats the $L$
levels as one block, so it can commit them in any order. Each paradigm starts from
its public pretrained checkpoint, so the backbone changes with the
paradigm. AR is T5 1.1 base~\cite{raffel2020t5}, an encoder--decoder
pretrained on C4. Its encoder reads \enquote{retrieve:} and the query; its
decoder attends to them and generates the $L$ levels and an end token.
MDLM~\cite{sahoo2024mdlm} and BD3-LM~\cite{arriola2025bd3lm} share one
DiT~\cite{peebles2023dit} architecture (12 layers, width 768, 12 heads, no
timestep input), each pretrained on OpenWebText; BD3-LM has checkpoints
at block sizes 4 and 8. AR has 260M
parameters and the diffusion models 182M: comparable, not matched.

\textbf{Conditioning on the query.} Retrieval needs $p_\theta(c \mid q)$.
T5 reads the query by design, but the released diffusion models only
model text and take no query. We condition MDLM as LLaDA does for
instruction tuning~\cite{nie2025llada}: the prompt stays clean, and only
the response is masked and scored. Here the query (at most 128 tokens) is
a clean prefix of the docid, in training and at inference. At noise level
$t \in [0,1]$, each docid level is masked independently:
\begin{equation}
  z_t^l = \begin{cases} \mathbf{m} & \text{with probability } t,\\
    c_l & \text{otherwise,} \end{cases} \qquad l = 1, \dots, L,
  \label{eq:forward}
\end{equation}
where $\mathbf{m}$ is the \texttt{[MASK]} token. MDLM attends in both
directions over $[q; z_t]$. Block diffusion keeps the query as clean first
blocks under BD3-LM's block-causal mask, with a clean stream $[q; c]$ and
a noisy stream $[q; z_t]$. Each noisy docid block attends to itself, to
the query and to the clean earlier docid blocks. The clean stream holds
the gold docid in training and the levels decoded so far at inference.

\label{sec:meth-train}
\textbf{Training.} The $L{\times}K$ docid tokens are added to each model's
input embedding and output layer, and all weights are fine-tuned to learn
$p_\theta(c \mid q)$ from (query, docid) pairs, with each paradigm's own
loss. Let $M(z) = \{l : z^l =
\mathbf{m}\}$ be the masked levels. Block diffusion splits the docid into $B = L/k$ blocks $c^b$, each with
its own noise level $t_b \sim \mathcal{U}(0, 1)$, and factorises
$\log p_\theta(c \mid q) = \sum_b \log p_\theta(c^b \mid c^{<b}, q)$.
Then
\begin{align}
  \mathcal{L}_{\text{AR}} &= -\textstyle\sum_{l=1}^{L}
    \log p_\theta(c_l \mid c_{<l}, q), \label{eq:ar}\\
  \mathcal{L}_{\text{diff}} &= \tfrac{1}{L}\textstyle\sum_{b=1}^{B}
    \mathbb{E}_{t_b}\, \mathbb{E}_{z^b_{t_b}}
    \textstyle\sum_{l \in M(z^b_{t_b})}
    -\log p_\theta(c_l \mid z^b_{t_b}, c^{<b}, q). \label{eq:bd}
\end{align}
With one block ($B{=}1$), $\mathcal{L}_{\text{diff}}$ trains MDLM.
MDLM and BD3-LM~\cite{sahoo2024mdlm,arriola2025bd3lm} weight each block's
term by $1/t_b$, which makes the loss a bound on the likelihood; we drop
this weight.

\label{sec:meth-decode}
\textbf{Decoding.} A trained model gives probabilities over codes, but
retrieval needs a ranked list of documents. AR bridges the two with beam
search (width 100) in a trie of corpus docids~\cite{tay2022dsi,wang2022nci},
stored as STATIC's sparse matrix~\cite{su2026static}. At each level, the
trie keeps only the codes that extend the current prefix to some corpus
docid. So every beam ends on a docid, and beams are ranked by summed
log-probability. Greedy decoding keeps one beam, with or without the
trie; without it, a code that is no docid counts as a miss.

Diffusion has no trie. We therefore test a ladder of decodings on the same
checkpoint, from sampling alone to scoring the whole corpus. Sampling
decodes block by block, with $S$ reverse
steps~\cite{austin2021d3pm,sahoo2024mdlm} per block. MDLM decodes the
whole docid as one block of $L$ levels, as it is trained ($S{=}16$); BD$k$
decodes blocks of $k$ levels ($S{=}k$). A reverse step
lowers the noise level from $t$ to $s < t$: the model predicts every
masked level, and each is revealed with probability $(t-s)/t$. A
stochastic sample draws the revealed codes from the prediction. Argmax
takes the most likely outcome instead. Staying masked is the most likely
outcome until a block's last step, so argmax commits a whole block at once.
For MDLM, argmax thus commits all $L$ levels together, from its prediction
on the fully masked docid.

A generated code $\hat c$ may not be a docid. \emph{Code matching} keeps
the 100 documents whose codes differ from $\hat c$ at the fewest levels,
ranked by that distance, ties in corpus order. The \emph{rerank} reorders
them by a per-level score from one pass on $z_1$, the fully masked docid:
\begin{equation}
  s(d) = \textstyle\sum_{l=1}^{L} \log p_\theta\big(c_l(d) \mid z_1, q\big).
  \label{eq:snap}
\end{equation}
\textbf{Generate-and-match}, our reference diffusion decoding, is argmax,
then code matching, then the rerank. \emph{Best-of-$N$} draws $N{=}8$
stochastic samples instead, pools the 100 nearest documents of each, and
reranks the pool by $s(d)$.

\textbf{One-pass scoring} skips generation. It ranks the whole corpus by
$s(d)$ and keeps the top 100. Generate-and-match uses the same score,
but only on the neighbours of one generated code. The pass on $z_1$ outputs an $L{\times}K$
table of log-probabilities, so scoring a document costs $L$ table
lookups. For MDLM, $s(d)$ is the $t{=}1$ term of its bound; for block
diffusion it is not, since that bound conditions each block on clean
earlier blocks. \emph{Chain-rule rescoring}
approximates the joint likelihood of the code on a shortlist: it reranks the top 20 of
one-pass scoring by $\sum_l \log p_\theta\big(c_l(d) \mid c_{<l}(d),
q\big)$, with one pass per candidate and level.

\section{Experimental Setup}
\label{sec:setup}

\label{sec:data}
\textbf{Data.} We use DDRO's two benchmarks, built with its
scripts\footnote{\url{https://github.com/kidist-amde/ddro}}~\cite{mekonnen2025ddro}.
NQ320K~\cite{kwiatkowski2019nq,wang2022nci} has 320k Wikipedia
query--document pairs with natural-language queries, documents
deduplicated by title, and the predefined training and validation splits.
Its \emph{unseen} documents are the gold of eval queries only; their
queries (22.4\%) are reported separately. MS300K is a subset of the
MS~MARCO document ranking dataset~\cite{bajaj2016msmarco} with 320k
documents and 360k training pairs; its eval queries are the official dev
queries whose gold is in the corpus, so it has no unseen documents.

\textbf{Identifiers.} Docids have $L{=}16$ levels of $K{=}512$ codes: 8,192
new tokens, each initialised at the mean pretrained row. RQ and PQ quantise
the GTR-T5-base embedding~\cite{ni2022gtr} ($D{=}768$). Random codes are
drawn per document with the same seed in both corpora, which also share
the random level permutation.

\label{sec:impl}
\textbf{Implementation.} Models train only on real training queries:
unlike DDRO, we map no document text and no generated queries to docids,
so a document that no training query maps to is never a target. Training
uses AdamW, gradient clipping at 1, dropout 0.1, bfloat16 and 16.4M
examples on one RTX 6000 Ada. AR uses a peak rate of $3{\cdot}10^{-3}$,
$\beta_2{=}0.95$, weight decay 0.01, linear decay and label smoothing 0.1;
diffusion uses $\beta_2{=}0.98$, weight decay $10^{-5}$, 3\% warmup and a
constant rate ($1.5{\cdot}10^{-2}$ for MDLM; $10^{-2}$ / $7{\cdot}10^{-3}$
for block diffusion on \NQ{} / \MSM{}), each the best of a short sweep on
RQ. We evaluate the final, not the EMA, weights. Code\codeurl{} and data
with all docids\dataurl{} are released.

\label{sec:meth-eval}
\textbf{Evaluation.} Hit@$k$, $k \in \{1, 10, 100\}$, is the share of
queries whose gold is in the top $k$. Documents that share a code cannot be
told apart, so hits are scored on the code. Intervals are paired 95\%
bootstrap intervals on a difference between two systems, resampling the
queries.

\textbf{Analyses.} A paradigm--identifier interaction counts only beyond
$\pm2$ Hit@1 points. \emph{Training exposure} groups eval queries by the
number of training queries of their gold. A \emph{miss}, a query with Hit@1${}=0$, takes the first
matching class: near-duplicate of the gold (same title, or character
5-gram Jaccard $\ge 0.8$), code with at most two wrong levels, topical
(TF-IDF cosine above the 99th percentile of gold--random pairs), or
unrelated. \emph{Gap closed} is the share of the Hit@1 gap between a
reference diffusion decoding and AR's beam that one-pass scoring recovers;
\emph{miss recovery}, the share of rank-1 misses whose gold stays among
the reranked candidates.

\section{Results}
\label{sec:results}

\begin{table}[t]
\footnotesize
  \centering
  \caption{AR leads or ties every learned-code column; on \NQ{}, random
    codes keep most of each paradigm's Hit@1. Hit@1 (\%); AR: trie beam,
    diffusion: generate-and-match. rev./shuf.: RQ with reversed / shuffled
    levels.
    Bold: best model per column, not a significance claim; grey: not converged. For
    reference, on the same benchmarks, DDRO~\cite{mekonnen2025ddro} reports
    (\NQ{} / \MSM{}): BM25 14.1 / 18.9, SEAL~\cite{bevilacqua2022seal} 29.3 / 27.6,
    NCI~\cite{wang2022nci} 32.7 / 29.5, dense ANCE~\cite{xiong2021ance}
    24.5 / 29.7, DDRO (PQ) 48.9 / 32.9.}
  \label{tab:arms}
  \tablefont
  \begin{tabular*}{\linewidth}{@{\extracolsep{\fill}}l*{10}{w{c}{\armscolw}}@{}}
    \toprule
    & \multicolumn{5}{c}{\NQ{}} & \multicolumn{5}{c}{\MSM{}} \\
    \cmidrule(lr){2-6}\cmidrule(lr){7-11}
    & \RQ{} & rev. & shuf. & \PQ{} & rand. & \RQ{} & rev. & shuf. & \PQ{} & rand. \\
    \midrule
    \pdot{arGray}\AR{} & \best{45.4} & \best{46.8} & \best{47.7} & \best{48.8} & \best{40.0} & \best{21.3} & \best{17.1} & \best{18.0} & \best{24.9} & \textcolor{gray}{0.3} \\
    \pdot{mdlmOrange}\MDLM{} & 41.1 & 40.6 & 40.3 & 44.7 & 34.3 & 18.1 & 15.6 & 16.6 & 16.8 & \textcolor{gray}{7.2} \\
    \pdot{bdeightBlue}\BDeight{} & 39.1 & 39.5 & 39.2 & 41.3 & 35.2 & 16.3 & \best{17.1} & 16.0 & 17.3 & \textcolor{gray}{6.2} \\
    \pdot{bdfourGreen}\BDfour{} & 37.5 & 37.5 & 37.5 & 36.8 & 32.9 & 15.8 & 14.7 & 14.2 & 14.0 & \textcolor{gray}{5.7} \\
    \bottomrule
  \end{tabular*}
\end{table}

\subsection{What Each Paradigm Learns}
\label{sec:training}

\begin{figure}[t]
  \centering
  \pgfplotstableread[col sep=comma, row sep=\\]{k,all_ar_pq,all_mdlm_pq,all_bd8_pq,all_bd4_pq,all_ar_rq,all_mdlm_rq,all_bd8_rq,all_bd4_rq,unseen_ar_pq,unseen_mdlm_pq,unseen_bd8_pq,unseen_bd4_pq,unseen_ar_rq,unseen_mdlm_rq,unseen_bd8_rq,unseen_bd4_rq\\
1,48.83,45.01,42.22,40.72,45.42,41.98,40.88,39.35,3.19,2.74,1.82,1.48,2.05,2.91,1.88,1.94\\
2,55.29,52.16,49.14,47.89,50.66,48.76,48.47,46.46,4.44,6.61,2.85,2.62,3.3,6.61,5.07,4.16\\
3,57.98,55.48,52.55,51.11,53.0,52.17,52.02,49.74,5.58,8.49,3.82,3.08,4.22,9.46,7.24,6.15\\
5,60.73,59.0,56.28,54.62,55.63,56.02,55.34,52.82,6.38,11.45,4.9,4.05,5.64,14.13,10.66,9.46\\
10,63.84,63.47,60.64,58.98,58.88,60.11,59.22,56.69,8.72,16.35,6.84,5.7,7.29,19.66,16.35,14.7\\
20,66.64,67.66,67.46,65.4,61.83,63.56,62.62,60.17,11.91,23.13,23.53,22.11,10.03,24.39,23.36,22.22\\
30,68.25,70.4,69.9,68.47,63.33,65.43,64.73,62.35,13.68,27.18,27.24,25.93,12.08,26.89,26.44,25.13\\
50,69.97,73.51,73.38,72.17,65.56,67.93,67.13,64.87,16.58,32.65,32.99,31.79,14.99,31.17,30.14,28.32\\
100,72.52,77.74,77.47,76.36,67.64,70.73,70.17,68.15,20.57,40.74,40.46,40.11,18.52,37.15,36.13,33.96\\
}\datadepth
\pgfplotstableread[col sep=comma, row sep=\\]{x,bin,n,ar_pq,mdlm_pq,bd8_pq,bd4_pq,ar_rand,mdlm_rand,bd8_rand,bd4_rand\\
0,0,1755,3.19,2.45,2.85,2.34,0.0,0.0,0.0,0.0\\
1,1,930,27.1,19.03,22.26,20.86,17.85,3.76,3.76,3.87\\
2,2-4,1543,54.18,45.88,45.3,42.19,38.56,24.3,27.03,27.61\\
3,5-12,1721,68.16,63.34,59.44,54.45,56.89,55.03,53.05,48.81\\
4,13+,1881,80.06,74.38,66.77,56.14,74.11,70.44,74.0,67.52\\
}\datamemorisation
\begin{tikzpicture}
  \begin{groupplot}[
    group style={group size=3 by 1, horizontal sep=0.85cm},
    figaxis, width=3.25cm, height=1.1cm,
  ]
    \nextgroupplot[xmin=-0.25, xmax=4.25,
      xtick={0,...,4}, xticklabels={0,1,2--4,5--12,13+},
      xlabel={(a) training queries of the gold}, ymin=0, ymax=85,
      ytick={0,40,80}, ylabel={Hit@$k$ (\%)},
      legend style={at={(1.76,1.04)}, anchor=south, legend columns=6,
        /tikz/every even column/.append style={column sep=5pt}}]
      \addplot[paradigmline=arGray] table[x=x, y=ar_pq]{\datamemorisation};
      \addlegendentry{AR}
      \addplot[paradigmline=mdlmOrange] table[x=x, y=mdlm_pq]{\datamemorisation};
      \addlegendentry{MDLM}
      \addplot[paradigmline=bdeightBlue] table[x=x, y=bd8_pq]{\datamemorisation};
      \addlegendentry{BD8}
      \addplot[paradigmline=bdfourGreen] table[x=x, y=bd4_pq]{\datamemorisation};
      \addlegendentry{BD4}
      \addlegendimage{draw=figInk, line width=0.9pt, mark=*, mark size=1.3pt, mark options={solid, fill=figInk}}
      \addlegendentry{PQ}
      \addlegendimage{draw=figInk, line width=0.9pt, densely dotted, mark=*, mark size=1.3pt, mark options={solid, fill=white}}
      \addlegendentry{random}
      \addplot[forget plot, paradigmline=arGray, densely dotted, mark options={solid, fill=white}] table[x=x, y=ar_rand]{\datamemorisation};
      \addplot[forget plot, paradigmline=mdlmOrange, densely dotted, mark options={solid, fill=white}] table[x=x, y=mdlm_rand]{\datamemorisation};
      \addplot[forget plot, paradigmline=bdeightBlue, densely dotted, mark options={solid, fill=white}] table[x=x, y=bd8_rand]{\datamemorisation};
      \addplot[forget plot, paradigmline=bdfourGreen, densely dotted, mark options={solid, fill=white}] table[x=x, y=bd4_rand]{\datamemorisation};
    \nextgroupplot[xmode=log,
      xmin=0.85, xmax=118, xtick={1,10,100}, xticklabels={1,10,100},
      xlabel={(b) $k$, all queries}, ymin=40, ymax=80, ytick={40,60,80}]
      \addplot[forget plot, paradigmline=arGray] table[x=k, y=all_ar_pq]{\datadepth};
      \addplot[forget plot, paradigmline=mdlmOrange] table[x=k, y=all_mdlm_pq]{\datadepth};
      \addplot[forget plot, paradigmline=bdeightBlue] table[x=k, y=all_bd8_pq]{\datadepth};
      \addplot[forget plot, paradigmline=bdfourGreen] table[x=k, y=all_bd4_pq]{\datadepth};
    \nextgroupplot[xmode=log,
      xmin=0.85, xmax=118, xtick={1,10,100}, xticklabels={1,10,100},
      xlabel={(c) $k$, unseen documents}, ymin=0, ymax=45, ytick={0,20,40}]
      \addplot[forget plot, paradigmline=arGray] table[x=k, y=unseen_ar_pq]{\datadepth};
      \addplot[forget plot, paradigmline=mdlmOrange] table[x=k, y=unseen_mdlm_pq]{\datadepth};
      \addplot[forget plot, paradigmline=bdeightBlue] table[x=k, y=unseen_bd8_pq]{\datadepth};
      \addplot[forget plot, paradigmline=bdfourGreen] table[x=k, y=unseen_bd4_pq]{\datadepth};
  \end{groupplot}
\end{tikzpicture}
  \caption{Documents with more training queries are retrieved more often;
    at depth, the best diffusion decoding reaches unseen documents more
    often than AR (\NQ{}). (a) Hit@1 by the gold's training queries (0:
    unseen); AR beam, diffusion generate-and-match. (b, c) Hit@$k$, PQ: AR
    beam; MDLM one-pass; BD8/BD4 chain-rule rescoring.}
  \label{fig:recall}
\end{figure}
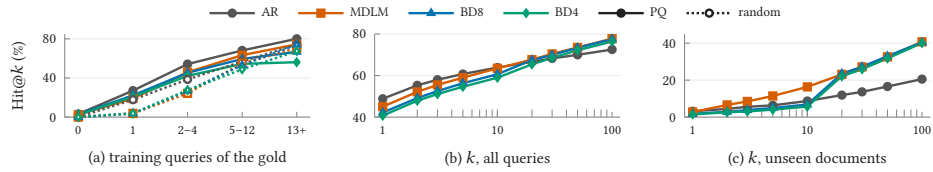

\textbf{AR leads or ties every learned-code cell} (\cref{tab:arms}), for
three possible reasons. First, AR predicts the first and last levels more accurately. On
\NQ{} eval queries, from the query alone, AR predicts the first level
correctly 61\% of the time, against 49--50\% for diffusion with every
level masked on PQ (RQ: 70 against 59--60\%); given the other 15 gold
levels, it predicts the last level correctly 79\% of the time, against
67--78\% (RQ: 82 against 54--58\%). Second, AR is trained on all 17
outputs (16 levels and the end token) of every example and masked diffusion on about half, so a fixed
example budget favours AR. Third, AR's backbone is larger and pretrained
differently: from random weights, AR scores 40.6 on \NQ{} PQ, 8.2 points
below its \cref{tab:arms} score and below MDLM's 44.7.

\textbf{Each paradigm needs its own training recipe.} Without weight
decay, AR training on \NQ{} can diverge; with 0.01, it converges. With
AR's recipe, every diffusion model we tried outputs a constant (MDLM-PQ
and BD8-PQ on \NQ{}, MDLM-RQ on \MSM{}: 44.7, 41.3 and 18.1 $\to$ 0.0).

\textbf{On \NQ{}, random codes keep most of each paradigm's Hit@1}
(\cref{tab:arms}, rand.): 83--90\% of RQ's, with no semantic meaning. The
models mostly memorise which docid answers which query, as known for
AR~\cite{pradeep2023scale,nicolas2026redsi}; this also holds for
diffusion. On \MSM{}, random codes keep less than half within the budget
(grey: not converged), so content-based codes learn faster.

\textbf{Documents with more training queries are retrieved more often}
(\cref{fig:recall}a). Grouped by training exposure (\cref{sec:setup}),
Hit@1 rises with the gold's number of training queries in every cell.
Content matters only when queries are scarce: with one training query,
random codes reach 4\% in every diffusion model, against 19--22\% for PQ;
above twelve, they catch up. The 1,755 eval queries with an unseen gold
get 2.3--3.2\% Hit@1 with PQ and none with random codes. AR's lead and the
PQ--RQ gap both come from seen documents: at each paradigm's best decoding,
AR leads MDLM by 4.8 points on seen documents (62.0 against 57.2, PQ; not
plotted) and
by 0.5 on unseen ones, and AR's PQ--RQ gap is 4.0 on seen documents.
Unseen documents appear deeper in the ranking (\cref{sec:generation}).

\textbf{On \NQ{}, level order matters little for diffusion} (\cref{tab:arms}, rev.,
shuf.). Reversing or shuffling the RQ levels moves diffusion's Hit@1 by
less than a point on \NQ{} and by $-2.5$ to $+0.8$ on \MSM{}. It raises
AR's by 1.4 and 2.3 on \NQ{} and costs it 4.2 and 3.3 on \MSM{}.

\subsection{What Diffusion Generates and How to Decode It}
\label{sec:generation}

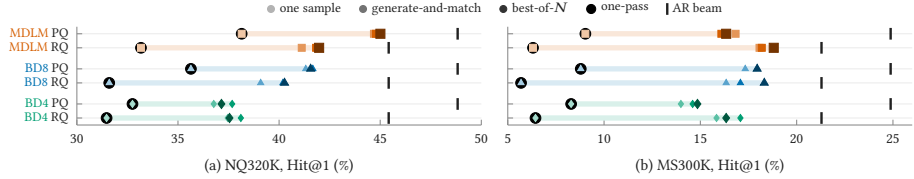
\begin{figure}[t]
  \centering
  \pgfplotstableread[col sep=comma, row sep=\\]{y,paradigm,identifier,stoch,std,bon,onepass,arbeam,lo,span\\
1,mdlm,pq,0.0903,0.1683,0.1609,0.1634,0.2488,0.0903,0.0780\\
2,mdlm,rq,0.0631,0.1807,0.1819,0.1881,0.2129,0.0631,0.1250\\
3.5,bd8,pq,0.0879,0.1733,0.1795,0.1795,0.2488,0.0879,0.0916\\
4.5,bd8,rq,0.0569,0.1634,0.1708,0.1832,0.2129,0.0569,0.1263\\
6,bd4,pq,0.0829,0.1399,0.1460,0.1485,0.2488,0.0829,0.0656\\
7,bd4,rq,0.0644,0.1584,0.1708,0.1634,0.2129,0.0644,0.1064\\
}\datadecodingdotsms
\pgfplotstableread[col sep=comma, row sep=\\]{y,paradigm,identifier,stoch,std,bon,onepass,arbeam,lo,span\\
1,mdlm,pq,0.3815,0.4469,0.4479,0.4501,0.4883,0.3815,0.0686\\
2,mdlm,rq,0.3317,0.4112,0.4183,0.4198,0.4542,0.3317,0.0881\\
3.5,bd8,pq,0.3564,0.4132,0.4167,0.4156,0.4883,0.3564,0.0603\\
4.5,bd8,rq,0.3160,0.3909,0.4019,0.4027,0.4542,0.3160,0.0867\\
6,bd4,pq,0.3275,0.3677,0.3768,0.3715,0.4883,0.3275,0.0493\\
7,bd4,rq,0.3147,0.3746,0.3811,0.3755,0.4542,0.3147,0.0664\\
}\datadecodingdotsnq
\begin{tikzpicture}
  \pgfplotsset{
    dec/.style 2 args={forget plot, only marks, pmark #1, mark size=1.3pt,
      draw=#1!#2, fill=#1!#2},
    onepass/.style={forget plot, only marks, pmark #1, mark size=1.7pt,
      draw=#1!55!black, fill=#1!55!black},
    arbeam/.style={forget plot, only marks, mark=|, mark size=2.2pt,
      mark options={draw=figInk, line width=0.9pt}},
    span/.style={forget plot, only marks, mark=none,
      error bars/.cd, x dir=plus, x explicit,
      error mark=none, error bar style={draw=#1!15, line width=2pt}},
    dotaxis/.style={figaxis, width=5.35cm, height=1.3cm,
      y dir=reverse, ymin=0.5, ymax=7.5,
      ytick={1,2,3.5,4.5,6,7}, ymajorgrids, xmajorgrids=false,
      percent x},
    legdec/.style={only marks, mark=*, mark size=1.3pt, draw=#1, fill=#1},
  }
  \newcommand\cellrows[3]{%
    \addplot[restrict y to domain=#3, span=#2] table[x=lo, y=y, x error=span]{#1};
    \addplot[dec={#2}{35}, restrict y to domain=#3] table[x=stoch, y=y]{#1};
    \addplot[dec={#2}{65}, restrict y to domain=#3] table[x=std, y=y]{#1};
    \addplot[dec={#2}{100}, restrict y to domain=#3] table[x=bon, y=y]{#1};
    \addplot[onepass=#2, restrict y to domain=#3] table[x=onepass, y=y]{#1};
  }
  \newcommand\allrows[1]{%
    \cellrows{#1}{mdlmOrange}{0.5:2.5}
    \cellrows{#1}{bdeightBlue}{3:5}
    \cellrows{#1}{bdfourGreen}{5.5:7.5}
    \addplot[arbeam] table[x=arbeam, y=y]{#1};
  }
  \begin{groupplot}[group style={group size=2 by 1, horizontal sep=0.35cm,
      yticklabels at=edge left}, dotaxis]
    \nextgroupplot[xlabel={(a) NQ320K, Hit@1 (\%)}, xmin=0.30, xmax=0.50,
      xtick={0.30,0.35,0.40,0.45,0.50},
      yticklabels={\textcolor{mdlmOrange}{MDLM} PQ, \textcolor{mdlmOrange}{MDLM} RQ,
        \textcolor{bdeightBlue}{BD8} PQ, \textcolor{bdeightBlue}{BD8} RQ,
        \textcolor{bdfourGreen}{BD4} PQ, \textcolor{bdfourGreen}{BD4} RQ},
      legend style={at={(1.04,1.0)}, anchor=south, legend columns=5,
        /tikz/every even column/.append style={column sep=5pt}}]
      \addlegendimage{legdec=black!30}\addlegendentry{one sample}
      \addlegendimage{legdec=black!55}\addlegendentry{generate-and-match}
      \addlegendimage{legdec=black!80}\addlegendentry{best-of-$N$}
      \addlegendimage{legdec=black!100, mark size=1.7pt}\addlegendentry{one-pass}
      \addlegendimage{only marks, mark=|, mark size=2.2pt, mark options={draw=figInk, line width=0.9pt}}\addlegendentry{AR beam}
      \allrows{\datadecodingdotsnq}
    \nextgroupplot[xlabel={(b) MS300K, Hit@1 (\%)}, xmin=0.05, xmax=0.26,
      xtick={0.05,0.10,0.15,0.20,0.25}]
      \allrows{\datadecodingdotsms}
  \end{groupplot}
\end{tikzpicture}
  \caption{Every diffusion model gains over one sample, and none reaches
    AR's beam. Each row: one diffusion model and identifier under four decodings (one sample, generate-and-match, best-of-$N$,
    one-pass), darker in that order (bar: their range). Black tick: AR's beam on the same
    identifier.}
  \label{fig:decoding}
\end{figure}

\begin{table}[t]
\footnotesize
  \centering
  \caption{One-pass scoring matches or beats generate-and-match in 11 of 12
    settings;
    chain-rule rescoring is mixed. $\Delta$ Hit@1 (points), first decoding
    minus second; grey: paired 95\% interval (\NQ{}; \MSM{} intervals not
    shown); bold: interval excludes 0.}
  \label{tab:snap}
  \tablefont
    \begin{tabular*}{\linewidth}{@{\extracolsep{\fill}}ll
      S[table-format=+1.1] l S[table-format=+1.1]
      S[table-format=+1.1] l S[table-format=+1.1]@{}}
    \toprule
    & & \multicolumn{3}{c}{One-pass vs generate-and-match} & \multicolumn{3}{c}{Chain-rule vs one-pass} \\
    \cmidrule(lr){3-5}\cmidrule(l){6-8}
    & & {\NQ{}} & & {\MSM{}} & {\NQ{}} & & {\MSM{}} \\
    \midrule
    \pdot{mdlmOrange}\MDLM{}     & \PQ{} & +0.3 & \ci{$-0.1$}{0.7} & -0.5 & \best{-3.1} & \ci{$-3.6$}{$-2.5$} & \best{-3.1} \\
                                 & \RQ{} & \best{+0.9} & \ci{0.5}{1.3} & +0.7 & \best{-0.6} & \ci{$-1.0$}{$-0.1$} & +1.1 \\
    \pdot{bdeightBlue}\BDeight{} & \PQ{} & +0.2 & \ci{0.0}{0.5} & +0.6 & +0.7 & \ci{$-0.1$}{1.5} & \best{-5.1} \\
                                 & \RQ{} & \best{+1.2} & \ci{0.8}{1.6} & +2.0 & \best{+0.6} & \ci{0.03}{1.2} & +1.1 \\
    \pdot{bdfourGreen}\BDfour{}  & \PQ{} & +0.4 & \ci{$-0.1$}{0.8} & +0.9 & \best{+3.6} & \ci{2.8}{4.4} & -0.5 \\
                                 & \RQ{} & +0.1 & \ci{$-0.4$}{0.6} & +0.5 & \best{+1.8} & \ci{1.1}{2.5} & +1.6 \\
    \bottomrule
  \end{tabular*}
\end{table}

\textbf{Raw generations rarely hit the gold.} Before code matching,
the generated code exactly matches the gold for only 14--21\% of \NQ{}
queries; after matching and rerank, Hit@1 is 37--45\%. Wrong generations
are far: 54--60\% differ at eight or more of the 16 levels. Matching
against the corpus, not denoising, carries much of the Hit@1.

\textbf{Decoding beyond one sample helps, and none reaches AR in Hit@1}
(\cref{fig:decoding}). For BD8-RQ on \NQ{}, Hit@1 goes from 31.6 (one
sample) to 35.4 (argmax), both with code matching, then 39.1 (generate-and-match),
40.2 (best-of-$N$) and 40.3 (one-pass). In every setting, argmax beats one sample and generate-and-match beats
argmax; best-of-$N$ moves Hit@1 by $-0.7$ to $+1.2$ from generate-and-match. From
one sample to the best decoding (chain rule included), Hit@1 rises by 6.6
to 13.7 points in every setting. AR's beam stays ahead of every row; the
closest plotted row is MDLM-RQ, 3.4 points behind on \NQ{} and 2.5 on \MSM{}. On
\NQ{}, AR's lead does not come from its beam (greedy is within a point);
on \MSM{}, the beam supplies 2.9 of its 6.9-point lead over the best diffusion
decoding on PQ. AR greedy
in the trie is also ahead of every diffusion decoding: by 3 to 7 points on \NQ{}
PQ, and by at least half a point elsewhere (over MDLM's chain-rule
rescoring on \MSM{} RQ). Without the trie, AR stays
ahead on \NQ{} (by 0.6 on RQ) but falls below MDLM
on \MSM{}.

\textbf{One-pass scoring nearly always matches or beats generate-and-match}
(\cref{tab:snap}, left). It is ahead in 11 of 12 settings, trails only for
MDLM-PQ on \MSM{} ($-0.5$), and its interval excludes zero in two of six
\NQ{} cells. Among misses, the share whose gold is missing from the
candidates falls by 19 to 31 points, so it mostly raises Hit@100, not
Hit@1. From generate-and-match, it closes 8 / 20\% (PQ / RQ) of MDLM's
gap to AR's beam on \NQ{} and $-6$ / 23\% on \MSM{}; from one sample,
64 / 72\% and 46 / 83\%. With every level masked, one-pass scoring sums
per-level marginals, so no score depends on a docid prefix: the model acts
as a per-level classifier.

\textbf{Chain-rule rescoring helps block diffusion on RQ and hurts MDLM on
PQ} (\cref{tab:snap}, right); elsewhere it is mixed (\NQ{} BD4-PQ
$+3.6$, \MSM{} BD8-PQ $-5.1$; MDLM-RQ $-0.6$ and $+1.1$). With every
level masked, block diffusion's per-level accuracy drops to about 0 at the first
level of every later block, a position it never predicts without the
earlier blocks in training; revealing them restores that signal. On PQ,
MDLM's accuracy is flat when all is masked (47--52\%), so there is no
lost signal for revealed levels to restore. On \MSM{}, only the two PQ losses exclude zero; on
\NQ{}, the split follows the training objective, not the identifier.

\textbf{At depth, the order reverses on \NQ{}} (\cref{fig:recall}b, c).
With one-pass scoring, MDLM ties AR's beam at Hit@10 on PQ (63.5 against
63.8) and leads on RQ (60.1 against 58.9; RQ not plotted). At Hit@100, it leads by 5.2
points on PQ and 3.1 on RQ, and reaches 41 against 21\% of unseen
documents. A 500-wide beam leaves AR's Hit@100 unchanged, so pruning is
not the cause. The depth gap matches what is proved for constrained beam
search~\cite{wu2025constrained}: ranking prefixes by marginal probability
keeps Hit@1 but loses depth, whereas one-pass scoring scores every docid.

\subsection{Does the Identifier Favour a Paradigm?}
\label{sec:identifier}

\begin{figure}[t]
  \centering
  \pgfplotstableread[col sep=comma, row sep=\\]{step,decode,nq_mdlm,nq_bd8,nq_bd4,ms_mdlm,ms_bd8,ms_bd4\\
0,stochastic,0.0498,0.0404,0.0128,0.0272,0.0310,0.0185\\
1,argmax,0.0302,0.0286,0.0042,-0.0099,0.0174,0.0124\\
2,+ rerank,0.0357,0.0223,-0.0069,-0.0124,0.0099,-0.0185\\
3,likelihood snap,0.0303,0.0129,-0.0040,-0.0247,-0.0037,-0.0149\\
4,chain rule,0.0050,0.0134,0.0137,-0.0669,-0.0656,-0.0359\\
}\dataidentifierdecodegap
\pgfplotstableread[col sep=comma, row sep=\\]{x,levels,mdlm_pq,bd8_pq,bd4_pq,mdlm_rq,bd8_rq,bd4_rq\\
0,1,26.6,30.5,27.2,11.7,11.9,12.3\\
1,2-4,68.3,71.3,69.0,29.2,30.8,28.5\\
2,5+,96.2,95.2,97.2,68.3,72.4,69.2\\
}\dataidentifierrecovery
\begin{tikzpicture}
  \begin{groupplot}[
    group style={group size=3 by 1, horizontal sep=0.8cm},
    figaxis, height=1.1cm,
  ]
    \nextgroupplot[width=3.15cm, xmin=-0.15, xmax=2.15,
      xtick={0,1,2}, xticklabels={1,2--4,5+}, ymin=0, ymax=100,
      ytick={0,50,100}, xlabel={(a) recovery by first wrong level},
      ylabel={miss recovery (\%)}]
      \addplot[forget plot, paradigmline=mdlmOrange] table[x=x, y=mdlm_pq]{\dataidentifierrecovery};
      \addplot[forget plot, paradigmline=bdeightBlue] table[x=x, y=bd8_pq]{\dataidentifierrecovery};
      \addplot[forget plot, paradigmline=bdfourGreen] table[x=x, y=bd4_pq]{\dataidentifierrecovery};
      \addplot[forget plot, paradigmline=mdlmOrange, dashed, mark options={solid, fill=white}] table[x=x, y=mdlm_rq]{\dataidentifierrecovery};
      \addplot[forget plot, paradigmline=bdeightBlue, dashed, mark options={solid, fill=white}] table[x=x, y=bd8_rq]{\dataidentifierrecovery};
      \addplot[forget plot, paradigmline=bdfourGreen, dashed, mark options={solid, fill=white}] table[x=x, y=bd4_rq]{\dataidentifierrecovery};
      \node[font=\tiny, text=figInk, anchor=south east] at (axis cs:1.05,76) {PQ};
      \node[font=\tiny, text=figInk, anchor=north west] at (axis cs:1.1,24) {RQ};
    \nextgroupplot[xlabel={(b) PQ $-$ RQ, NQ320K}, width=2.95cm, xmin=-0.3, xmax=4.3,
      xtick={0,...,4}, xticklabels={s,a,g,l,r},
      ymin=-0.075, ymax=0.055, ytick={-0.06,-0.03,0,0.03},
      ymajorgrids=false, percent y]
      \draw[draw=figMuted, line width=0.4pt] (axis cs:-0.3,0) -- (axis cs:4.3,0);
      \addplot[forget plot, paradigmline=mdlmOrange] table[x=step, y=nq_mdlm]{\dataidentifierdecodegap};
      \addplot[forget plot, paradigmline=bdeightBlue] table[x=step, y=nq_bd8]{\dataidentifierdecodegap};
      \addplot[forget plot, paradigmline=bdfourGreen] table[x=step, y=nq_bd4]{\dataidentifierdecodegap};
    \nextgroupplot[xlabel={(c) PQ $-$ RQ, MS300K}, width=2.95cm, xmin=-0.3, xmax=4.3,
      xtick={0,...,4}, xticklabels={s,a,g,l,r},
      ymin=-0.075, ymax=0.055, ytick={-0.06,-0.03,0,0.03},
      ymajorgrids=false, percent y]
      \draw[draw=figMuted, line width=0.4pt] (axis cs:-0.3,0) -- (axis cs:4.3,0);
      \addplot[forget plot, paradigmline=mdlmOrange] table[x=step, y=ms_mdlm]{\dataidentifierdecodegap};
      \addplot[forget plot, paradigmline=bdeightBlue] table[x=step, y=ms_bd8]{\dataidentifierdecodegap};
      \addplot[forget plot, paradigmline=bdfourGreen] table[x=step, y=ms_bd4]{\dataidentifierdecodegap};
  \end{groupplot}
\end{tikzpicture}
  \caption{PQ keeps the gold within reach after a miss, and its lead
    depends on the decoding. (a) Miss recovery (\%, \NQ{},
    generate-and-match), by the first wrong level.
    (b, c) Hit@1(PQ) $-$ Hit@1(RQ) in points for each diffusion model, by
    decoding (s: one sample;
    a: argmax; g: generate-and-match; l: one-pass; r: chain rule).}
  \label{fig:identifier}
\end{figure}
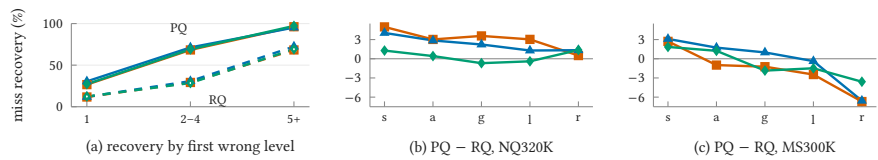

\textbf{On \NQ{}, the identifier does not favour diffusion over AR}
(\cref{tab:arms}). If PQ-like codes suit parallel decoders
(\cref{sec:related}), PQ's lead should grow with the number of levels a model commits at once. On
\NQ{}, PQ leads RQ by 3.4 points under AR, 3.6 under MDLM and 2.2 under
BD8. Under BD4 the two tie ($-0.7$). Within diffusion, PQ's lead grows
with the levels committed at once (BD4 $-0.7$, BD8 2.2, MDLM 3.6), but AR,
one level at a time, matches MDLM (3.4). AR's gap minus the mean
diffusion gap is $+1.7$ points with generate-and-match, inside our $\pm2$-point
threshold, but 2.1 with one-pass scoring and 2.3 with chain-rule rescoring
(\cref{fig:identifier}b). At each model's best decoding, it is 1.5, inside
the threshold. On \MSM{}, the point
estimates go the other way: AR's beam favours PQ by 3.6 points, and
diffusion with chain-rule rescoring favours RQ by 3.6 to 6.7, an
interaction of 4.3 to 9.2 points across decodings, a hint against the
parallel-decoder hypothesis.

\textbf{PQ's lead depends on the decoding} (\cref{fig:identifier}b, c).
With one stochastic sample, PQ leads in all six diffusion cells, by 1.3 to
5.0 points. Better decoding shrinks this lead on \NQ{} for MDLM and BD8, though not
steadily (MDLM: 5.0, 3.0, 3.6, then 3.0 with one-pass scoring). On \MSM{},
every diffusion model favours RQ with one-pass scoring and with chain-rule
rescoring. The rerank of the code-matching
candidates gains more on RQ in all four block-diffusion cells. A
one-sample comparison thus overstates PQ for diffusion.

\textbf{After a miss, the gold PQ document stays within reach; the gold
RQ one does not} (\cref{fig:identifier}a). We group misses by the first
level where the generation leaves the gold. When the generation leaves the
gold in its first four levels, the gold PQ document stays among the
candidates about 2.3 times as often as the gold RQ one; later, 1.3 to 1.4
times. The paradigms differ by only a few points. Recovery is thus a
property of the identifier, not of the paradigm. A near miss on PQ keeps
the other levels right, because they are independent. RQ's later levels
are small residual corrections, and diffusion's accuracy on them does not
rise with context (\cref{sec:training}). PQ leads although far more documents share a PQ code (1.4 against 0.1\%). Scored on the document, its lead
shrinks but keeps its sign (AR: $+3.4 \to +1.9$).

\subsection{Where Misses Land}
\label{sec:errors}

\begin{figure}[t]
  \centering
  \pgfplotstableread[col sep=comma, row sep=\\]{x,label,identifier,paradigm,decode,near_dup,code_near,topical,far\\
0.0,beam,pq,ar,beam,2.97,4.45,73.64,18.95\\
1.4,s,pq,mdlm,stochastic,1.50,3.67,65.85,28.98\\
2.4,c,pq,mdlm,canonical,2.05,5.87,73.61,18.48\\
3.4,l,pq,mdlm,A1,2.37,5.77,75.15,16.72\\
4.8,s,pq,bd8,stochastic,1.09,4.07,63.36,31.48\\
5.8,c,pq,bd8,canonical,2.10,5.69,75.90,16.32\\
6.8,l,pq,bd8,top20,2.13,7.24,71.88,18.75\\
8.2,s,pq,bd4,stochastic,1.48,3.51,63.83,31.17\\
9.2,c,pq,bd4,canonical,1.87,5.04,74.68,18.42\\
10.2,l,pq,bd4,top20,2.31,5.92,72.98,18.79\\
11.6,beam,rq,ar,beam,1.73,0.00,70.91,27.36\\
13.0,s,rq,mdlm,stochastic,0.66,0.13,27.87,71.33\\
14.0,c,rq,mdlm,canonical,1.95,0.15,74.32,23.57\\
15.0,l,rq,mdlm,A1,2.29,0.15,78.51,19.05\\
16.4,s,rq,bd8,stochastic,0.92,0.00,25.46,73.62\\
17.4,c,rq,bd8,canonical,2.07,0.30,74.85,22.78\\
18.4,l,rq,bd8,top20,2.30,0.31,77.88,19.51\\
19.8,s,rq,bd4,stochastic,1.19,0.00,28.04,70.77\\
20.8,c,rq,bd4,canonical,2.21,0.15,74.26,23.38\\
21.8,l,rq,bd4,top20,2.26,0.30,76.62,20.81\\
23.2,beam,random_v2,ar,beam,0.25,0.00,5.71,94.04\\
24.6,s,random_v2,mdlm,stochastic,0.25,0.00,4.39,95.36\\
25.6,c,random_v2,mdlm,canonical,0.80,0.00,24.80,74.40\\
26.6,l,random_v2,mdlm,A1,0.83,0.00,44.28,54.90\\
28.0,s,random_v2,bd8,stochastic,0.25,0.00,2.86,96.89\\
29.0,c,random_v2,bd8,canonical,1.58,0.00,20.05,78.36\\
30.0,l,random_v2,bd8,top20,0.94,0.00,27.04,72.02\\
31.4,s,random_v2,bd4,stochastic,0.37,0.00,2.86,96.77\\
32.4,c,random_v2,bd4,canonical,0.66,0.00,14.44,84.91\\
33.4,l,random_v2,bd4,top20,0.53,0.00,19.37,80.11\\
}\dataerrorscausesdecodesms
\pgfplotstableread[col sep=comma, row sep=\\]{x,label,identifier,paradigm,decode,near_dup,code_near,topical,far\\
0.0,beam,pq,ar,beam,0.77,1.62,52.16,45.45\\
1.4,s,pq,mdlm,stochastic,0.56,2.48,48.59,48.38\\
2.4,c,pq,mdlm,canonical,0.68,2.99,53.68,42.65\\
3.4,l,pq,mdlm,A1,0.72,3.07,55.60,40.62\\
4.8,s,pq,bd8,stochastic,0.58,2.66,46.60,50.17\\
5.8,c,pq,bd8,canonical,0.54,3.24,54.19,42.02\\
6.8,l,pq,bd8,top20,0.77,2.76,55.24,41.22\\
8.2,s,pq,bd4,stochastic,0.44,1.84,45.84,51.88\\
9.2,c,pq,bd4,canonical,0.53,2.93,52.17,44.37\\
10.2,l,pq,bd4,top20,0.67,2.67,54.24,42.42\\
11.6,beam,rq,ar,beam,1.12,0.12,43.92,54.84\\
13.0,s,rq,mdlm,stochastic,0.69,0.13,27.48,71.70\\
14.0,c,rq,mdlm,canonical,0.74,0.11,47.66,51.50\\
15.0,l,rq,mdlm,A1,0.88,0.11,50.12,48.89\\
16.4,s,rq,bd8,stochastic,0.41,0.06,25.60,73.94\\
17.4,c,rq,bd8,canonical,0.65,0.10,46.82,52.42\\
18.4,l,rq,bd8,top20,0.71,0.09,50.77,48.43\\
19.8,s,rq,bd4,stochastic,0.52,0.06,26.85,72.57\\
20.8,c,rq,bd4,canonical,0.67,0.08,46.11,53.13\\
21.8,l,rq,bd4,top20,0.82,0.06,50.98,48.14\\
23.2,beam,random_v2,ar,beam,1.30,0.00,38.25,60.46\\
24.6,s,random_v2,mdlm,stochastic,nan,nan,nan,nan\\
25.6,c,random_v2,mdlm,canonical,0.93,0.00,38.48,60.59\\
26.6,l,random_v2,mdlm,A1,1.01,0.00,44.79,54.20\\
28.0,s,random_v2,bd8,stochastic,nan,nan,nan,nan\\
29.0,c,random_v2,bd8,canonical,1.01,0.00,35.58,63.41\\
30.0,l,random_v2,bd8,top20,0.84,0.00,39.36,59.79\\
31.4,s,random_v2,bd4,stochastic,nan,nan,nan,nan\\
32.4,c,random_v2,bd4,canonical,0.82,0.00,27.81,71.38\\
33.4,l,random_v2,bd4,top20,0.85,0.00,33.42,65.73\\
}\dataerrorscausesdecodesnq
\definecolor{causeNear}{HTML}{882255}
\definecolor{causeCode}{HTML}{CC79A7}
\definecolor{causeTopic}{HTML}{C9A227}
\definecolor{causeFar}{HTML}{5BA3D9}
\begin{tikzpicture}
  \pgfplotsset{
    causeaxis/.style={width=10.8cm, height=1.05cm, ybar stacked,
      bar width=0.65, xmin=-0.7, xmax=33.9, ymin=0, ymax=100,
      ytick={0,50,100}, yticklabels={0,50,100\%}, ylabel style={align=center, font=\scriptsize, rotate=-90},
      xtick={0,1.4,2.4,3.4,4.8,5.8,6.8,8.2,9.2,10.2,
             11.6,13,14,15,16.4,17.4,18.4,19.8,20.8,21.8,
             23.2,24.6,25.6,26.6,28,29,30,31.4,32.4,33.4},
      xticklabels={beam,s,g,l,s,g,r,s,g,r,beam,s,g,l,s,g,r,s,g,r,beam,s,g,l,s,g,r,s,g,r},
      x tick label style={font=\tiny, inner sep=1pt},
      xmajorgrids=false, axis on top, clip=false},
    causebar/.style={draw=none, fill=#1},
  }
  \begin{groupplot}[group style={group size=1 by 2, vertical sep=0.3cm}, figaxis]
    \nextgroupplot[causeaxis, xticklabels={}, ylabel={(a)},
      legend style={at={(0.5,1.04)}, anchor=south, legend columns=4,
        /tikz/every even column/.append style={column sep=6pt}}]
      \addplot[causebar=causeNear] table[x=x, y=near_dup]{\dataerrorscausesdecodesnq};
      \addlegendentry{near-duplicate}
      \addplot[causebar=causeCode] table[x=x, y=code_near]{\dataerrorscausesdecodesnq};
      \addlegendentry{code within 2 levels}
      \addplot[causebar=causeTopic] table[x=x, y=topical]{\dataerrorscausesdecodesnq};
      \addlegendentry{same topic}
      \addplot[causebar=causeFar] table[x=x, y=far]{\dataerrorscausesdecodesnq};
      \addlegendentry{unrelated}
    \nextgroupplot[causeaxis, ylabel={(b)}]
      \addplot[forget plot, causebar=causeNear] table[x=x, y=near_dup]{\dataerrorscausesdecodesms};
      \addplot[forget plot, causebar=causeCode] table[x=x, y=code_near]{\dataerrorscausesdecodesms};
      \addplot[forget plot, causebar=causeTopic] table[x=x, y=topical]{\dataerrorscausesdecodesms};
      \addplot[forget plot, causebar=causeFar] table[x=x, y=far]{\dataerrorscausesdecodesms};
      \node[font=\tiny, text=figInk, anchor=north, yshift=-7pt] at (axis cs:0.0,0) {AR};
      \node[font=\tiny, text=figInk, anchor=north, yshift=-7pt] at (axis cs:2.4,0) {MDLM};
      \node[font=\tiny, text=figInk, anchor=north, yshift=-7pt] at (axis cs:5.8,0) {BD8};
      \node[font=\tiny, text=figInk, anchor=north, yshift=-7pt] at (axis cs:9.2,0) {BD4};
      \node[font=\scriptsize, text=figInk, anchor=north, yshift=-14pt] at (axis cs:5.1,0) {PQ};
      \node[font=\tiny, text=figInk, anchor=north, yshift=-7pt] at (axis cs:11.6,0) {AR};
      \node[font=\tiny, text=figInk, anchor=north, yshift=-7pt] at (axis cs:14.0,0) {MDLM};
      \node[font=\tiny, text=figInk, anchor=north, yshift=-7pt] at (axis cs:17.4,0) {BD8};
      \node[font=\tiny, text=figInk, anchor=north, yshift=-7pt] at (axis cs:20.8,0) {BD4};
      \node[font=\scriptsize, text=figInk, anchor=north, yshift=-14pt] at (axis cs:16.7,0) {RQ};
      \node[font=\tiny, text=figInk, anchor=north, yshift=-7pt] at (axis cs:23.2,0) {AR};
      \node[font=\tiny, text=figInk, anchor=north, yshift=-7pt] at (axis cs:25.6,0) {MDLM};
      \node[font=\tiny, text=figInk, anchor=north, yshift=-7pt] at (axis cs:29.0,0) {BD8};
      \node[font=\tiny, text=figInk, anchor=north, yshift=-7pt] at (axis cs:32.4,0) {BD4};
      \node[font=\scriptsize, text=figInk, anchor=north, yshift=-14pt] at (axis cs:28.3,0) {random};
  \end{groupplot}
\end{tikzpicture}
  \caption{With learned codes, the paradigm does not change where a miss
    lands. Rank-1 misses (\%) by type of wrong top-1 document; (a) \NQ{},
    (b) \MSM{}. beam: AR's trie beam; s: one sample; g: generate-and-match;
    l: one-pass scoring (MDLM); r: chain-rule rescoring (BD8, BD4).}
  \label{fig:errors}
\end{figure}
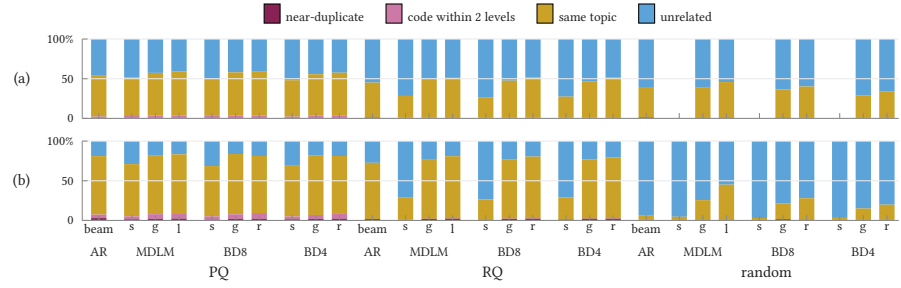

\textbf{With learned codes, the paradigm does not change where a miss
lands; the identifier, the corpus and the decoding do} (\cref{fig:errors}).
With PQ or RQ codes, one corpus and one decoding, the four paradigms share
one profile of causes. On topical misses, AR's beam is within 4
points of diffusion generate-and-match. The paradigm changes how many
queries miss, not what a miss returns. Under RQ, one stochastic sample
sends 71--74\% of misses to unrelated documents on both corpora. Generate-and-match,
one-pass scoring and chain-rule rescoring bring most of
them back on topic on \MSM{} (74--79\%, \cref{fig:errors}b); on \NQ{}, RQ
misses still split about evenly (\cref{fig:errors}a). Stochastic PQ misses
are already topical for 46--49\% on \NQ{} and 63--66\% on \MSM{}. Near-duplicates and near
codes stay at or below 10\%. Random-code misses are mostly unrelated
(54--97\%).
Across cells, misses only partly overlap: an oracle over the eight learned-code cells
(AR's beam, diffusion's generate-and-match) reaches 62.6\% Hit@1 on \NQ{},
against 48.8\% for the best cell (\MSM{}: 49.1
against 24.9).

\subsection{A Recipe for Diffusion GR}
\label{sec:recipe}

\textbf{The results give a recipe for diffusion retrievers.} First, train
diffusion with its own recipe (\cref{sec:training}). Second, decode with one-pass scoring. It matches or
beats generate-and-match in 11 of 12 diffusion settings
(\cref{tab:snap}), and on \NQ{}, MDLM's one-pass scoring reaches deeper
into the ranking than AR's beam. Third, for block diffusion on RQ, add chain-rule rescoring on the top 20
($+0.6$ to $+1.8$ on both corpora; it costs MDLM 3.1 on PQ). A paradigm comparison should report up to four numbers per paradigm:
the raw exact-match rate, Hit@1 at the paradigm's best decoding, and, where the corpus has unseen documents, Hit@1 on seen and unseen
ones separately.

\section{Limitations and Conclusion}
\label{sec:discussion}
\label{sec:conclusion}

\label{sec:limitations}
\textbf{Limitations.} Four design choices bound our
results. The first is the models. The paradigms do not share a
backbone: AR's T5 is pretrained on C4, about twenty times more text than
the diffusion models' OpenWebText, and since AR trained from random
weights falls below MDLM (\cref{sec:training}), part of AR's lead may
come from pretraining. Sizes are also only comparable, not matched
(\cref{sec:method}), and we keep them below 500M parameters because a
retriever over a few hundred thousand documents should stay small; as
larger models may generate better, whether the ranking holds at scale
stays open. The second is training. AR's recipe collapses diffusion
(\cref{sec:training}), so each paradigm trains at its own, which leaves
recipe and paradigm partly entangled. The third is the data and the docids. RQ3 covers
one code shape and numeric docids, and RQ1 and RQ3 rest on \NQ{}, because
on \MSM{} random codes do not converge within the budget and the identifier
gaps change sign across decodings. The last is the decoding. One-pass scoring is compared with
generative decodings only, not with a non-generative per-level
classifier, so how much of its gain needs diffusion is open, and its cost grows with
the corpus; why chain-rule rescoring costs MDLM 3.1 points on PQ is also
open.

\textbf{Conclusion.} In this paper, we crossed paradigm, identifier and
decoding on the same trained models to ask why diffusion retrievers trail
autoregressive ones. First, diffusion models memorise like
AR~\cite{pradeep2023scale}, contrary to the intuition behind semantic
docids~\cite{tay2022dsi,rajput2023tiger}:
on \NQ{}, random codes without meaning keep 83--90\% of RQ's Hit@1 in
every paradigm, though on \MSM{} they do not converge within the budget (RQ1), and
the identifier separates AR from diffusion by 1.5 points at each model's
best decoding on \NQ{}, inside our threshold (RQ3). Second, decoding alone moves diffusion Hit@1 by 6.6 to 13.7
points. The generated docid is right for only 14--21\% of \NQ{} queries,
yet diffusion needs no iterative denoising to retrieve: one forward pass
on a fully masked docid matches or beats generate-and-match in 11 of 12
settings and, on \NQ{}, reaches deeper than AR's beam (RQ2). What remains is
AR's lead in Hit@1, which on \NQ{} comes from the model rather than its
beam and is entangled with pretraining. What reads as a paradigm gap is
thus partly a decoding gap: a paradigm comparison is fair only at each
paradigm's own recipe and best decoding (\cref{sec:recipe}).
Our findings also open two directions. Since one forward pass yields an $L{\times}K$ table of code scores,
searching that table without scanning the corpus is a natural next step. Since every paradigm memorises, future work should test whether
training that reaches documents without a training query changes the
paradigm ranking.

\bibliographystyle{ACM-Reference-Format}
\bibliography{references}

\onecolumn
\let\figure\colfigure \let\endfigure\endcolfigure
\renewcommand{\tablefont}{\footnotesize}
\appendix

\section{Data, Identifiers and Settings}
\begin{table}[!htbp]
  \centering
  \caption{Settings of every grid cell. Learning rates come from a sweep on
    RQ codes and are reused for PQ and random codes. Every cell: mixed
    bfloat16, dropout 0.1, one RTX 6000 Ada; final weights evaluated (MDLM's
    EMA unused).}
  \label{tab:settings}
  \tablefont
  \setlength{\tabcolsep}{3pt}
  \begin{tabular*}{\linewidth}{@{\extracolsep{\fill}}l>{\raggedright\arraybackslash}p{0.26\linewidth}>{\raggedright\arraybackslash}p{0.47\linewidth}@{}}
    \toprule
    & {\AR{}} & {\MDLM{}, \BDeight{}, \BDfour{}} \\
    \midrule
    Initialisation & T5 1.1 base (C4) & DiT 12 layers, width 768, 12 heads, OpenWebText checkpoint (BD: one per block size) \\
    Size, time & 260M; 7 h & 182M; MDLM 4.5 h, BD 9.2 h \\
    Input & \enquote{retrieve:} + query; 16 codes + end token & GPT-2 tokens; query ($\le$128 tokens, never masked) + 16 codes \\
    Loss & cross-entropy, label smoothing 0.1 & cross-entropy on masked codes, unweighted; $t \sim \mathcal{U}(0,1)$ per example (BD: per block) \\
    Optimizer & AdamW, peak lr $3{\cdot}10^{-3}$, $\beta_2{=}0.95$, weight decay 0.01, gradient clip 1; linear decay, no warmup & AdamW, peak lr $1.5{\cdot}10^{-2}$ (MDLM), $10^{-2}$ / $7{\cdot}10^{-3}$ (BD, \NQ{} / \MSM{}), $\beta_2{=}0.98$, weight decay $10^{-5}$, gradient clip 1; 3\% linear warmup, then constant \\
    Budget & 64K steps $\times$ 256 & MDLM 32K $\times$ 512, BD 64K $\times$ 256; 16.4M examples in every cell \\
    Decoding & beam 100 in the trie & generate-and-match: argmax over 1 block $\times$ 16 steps (MDLM), 2 blocks $\times$ 8 (BD8), 4 blocks $\times$ 4 (BD4), code matching, rerank \\
    \bottomrule
  \end{tabular*}
\end{table}

\begin{table}[!htbp]
  \centering
  \caption{Corpora. Unseen: documents that are gold for no training query.
    Our \MSM{} is built with the DDRO script (title and content, truncated to
    4,000 characters), $\sim$10\% of \MSM{} Document; its eval set is DDRO's
    dev set, not the standard 5,193-query dev set.}
  \label{tab:data}
  \setlength{\tabcolsep}{4pt}
  \scriptsize
  \begin{tabular}{@{}lrr@{}}
    \toprule
    & \NQ{} & \MSM{} \\
    \midrule
    Documents & 109,739 & 319,927 \\
    Training queries & 307,373 & 367,013 \\
    Documents with a training query & 108,026 & 319,927 \\
    Eval queries & 7,830 & 808 \\
    Unseen documents / eval queries on them & 1,713 / 1,755 (22.4\%) & 0 / 0 \\
    \midrule
    Training examples per cell & 16M & 16M \\
    \quad per document & 146 & 50 \\
    \bottomrule
  \end{tabular}
\end{table}

\begin{table}[!htbp]
  \centering
  \caption{What the identifiers encode (16$\times$512). Levels 1-based, level 1 = coarsest; PQ is
    $\sim$8.96 bits at every level. Summed entropy: sum of per-level
    entropies; identifier entropy: bits that tell documents apart (at most
    $\log_2 N$ = 16.74 / 18.29). Collision rate: excess documents
    ($N-$distinct codes)$/N$. U@$k$: mean of 1/(size of a document's
    $k$-prefix group), corpus / gold documents. $^\ast$: last level only. \MSM{} RQ: 499 live codes at level 1, levels 3--5 carry $\sim$4.8 bits.
    }
  \label{tab:codebook}
  \setlength{\tabcolsep}{3pt}
  \scriptsize
  \begin{adjustbox}{max width=\linewidth}
\begin{tabular}{@{}lrr@{\hskip 8pt}rr@{}}
    \toprule
    & \multicolumn{2}{c}{\NQ{}} & \multicolumn{2}{c}{\MSM{}} \\
    \cmidrule(lr){2-3}\cmidrule(l){4-5}
    & \RQ{} & \PQ{} & \RQ{} & \PQ{} \\
    \midrule
    Summed per-level entropy (bits)  & 60.5  & 143.4  & 65.7  & 143.5 \\
    Identifier entropy (bits)  & 16.74  & 16.69  & 18.22  & 18.17 \\
    Level-1 entropy (bits)  & 7.86  & $\sim$8.96  & 7.95  & $\sim$8.96 \\
    Level-2 entropy (bits)  & 0.28  & $\sim$8.96  & 1.241  & $\sim$8.96 \\
    Documents sharing the top level-2 code  & 98.2\%  & 0.3--0.5\%  & 51.4\%  & 0.3--0.5\% \\
    Tail-level live codes / bits  & 28--41 / $\sim$3.8  & 512 / 8.96  & 32 / 3.97$^\ast$  & 512 / 8.97$^\ast$ \\
    Collision rate (excess documents)  & 0.091\% (100)  & 1.375\% (1,509)  & 1.925\% (6,160)  & 2.916\% (9,330) \\
    Largest collision bucket  & 5  & 75  & 512  & 737 \\
    U@4, corpus / gold  & .27 / .28  & .88 / .85  & .53 / .50  & .81 / .66 \\
    U@16, corpus / gold  & .999 / .998  & .986 / .979  & .981 / .958  & .971 / .937 \\
    Trie max branching  & 24  & 153  & 34  & 304 \\
    \bottomrule
  \end{tabular}
\end{adjustbox}
\end{table}

\clearpage

\section{Query Conditioning}
\begin{figure}[!htbp]
  \centering
  \begin{tikzpicture}[x=1cm, y=1cm, font=\scriptsize]
  \def\s{0.17}
  \tikzset{
    qcell/.style={fill=#1!35, draw=white, line width=0.2pt},
    off/.style={fill=figGrid, draw=white, line width=0.2pt},
    dcell/.style={fill=#1, draw=white, line width=0.2pt},
    grp/.style={font=\tiny, text=figInk},
    ttl/.style={font=\scriptsize, text=figInk, anchor=south west},
  }

  \begin{scope}[shift={(0,0)}]
    \node[ttl] at (0,0.05) {(a) AR (T5)};
    \foreach \i in {1,...,4}{
      \foreach \j in {1,...,8}{
        \pgfmathtruncatemacro{\att}{(\j<=4) || (\j-4<=\i)}
        \pgfmathtruncatemacro{\isq}{\j<=4}
        \ifnum\att=1
          \ifnum\isq=1 \path[qcell=arGray] (\j*\s-\s,-\i*\s) rectangle ++(\s,\s);
          \else \path[dcell=arGray] (\j*\s-\s,-\i*\s) rectangle ++(\s,\s); \fi
        \else \path[off] (\j*\s-\s,-\i*\s) rectangle ++(\s,\s); \fi
      }
    }
    \node[grp, anchor=north] at (2*\s,-4*\s-0.02) {$q$};
    \node[grp, anchor=north] at (6*\s,-4*\s-0.02) {$c$};
    \node[grp, anchor=east] at (-0.02,-2*\s) {$c$};
    \node[grp, anchor=north west, align=left, text width=2.0cm] at (0,-4*\s-0.3)
      {cross-attends to $q$;\\causal over $c$};
  \end{scope}

  \begin{scope}[shift={(2.25,0)}]
    \node[ttl] at (0,0.05) {(b) MDLM};
    \foreach \i in {1,...,8}{
      \foreach \j in {1,...,8}{
        \pgfmathtruncatemacro{\isq}{\j<=4}
        \ifnum\isq=1 \path[qcell=mdlmOrange] (\j*\s-\s,-\i*\s) rectangle ++(\s,\s);
        \else \path[dcell=mdlmOrange] (\j*\s-\s,-\i*\s) rectangle ++(\s,\s); \fi
      }
    }
    \node[grp, anchor=north] at (2*\s,-8*\s-0.02) {$q$};
    \node[grp, anchor=north] at (6*\s,-8*\s-0.02) {$z_t$};
    \node[grp, anchor=east] at (-0.02,-2*\s) {$q$};
    \node[grp, anchor=east] at (-0.02,-6*\s) {$z_t$};
  \end{scope}

  \begin{scope}[shift={(4.35,0)}]
    \node[ttl] at (0,0.05) {(c) block diffusion (BD3-LM)};
    \foreach \i in {1,...,16}{
      \foreach \j in {1,...,16}{
        \pgfmathtruncatemacro{\ni}{\i<=8}
        \pgfmathtruncatemacro{\nj}{\j<=8}
        \pgfmathtruncatemacro{\bi}{floor((\i-1-8*(1-\ni))/2)}
        \pgfmathtruncatemacro{\bj}{floor((\j-1-8*(1-\nj))/2)}
        \pgfmathtruncatemacro{\att}{(\ni==1 && \nj==1 && \bi==\bj)
          || (\ni==1 && \nj==0 && \bj<\bi) || (\ni==0 && \nj==0 && \bj<=\bi)}
        \pgfmathtruncatemacro{\isq}{\bj<=1}
        \ifnum\att=1
          \ifnum\isq=1 \path[qcell=bdeightBlue] (\j*\s-\s,-\i*\s) rectangle ++(\s,\s);
          \else \path[dcell=bdeightBlue] (\j*\s-\s,-\i*\s) rectangle ++(\s,\s); \fi
        \else \path[off] (\j*\s-\s,-\i*\s) rectangle ++(\s,\s); \fi
      }
    }
    \draw[figInk, line width=0.5pt] (8*\s,0) -- (8*\s,-16*\s);
    \draw[figInk, line width=0.5pt] (0,-8*\s) -- (16*\s,-8*\s);
    \node[grp, anchor=north] at (4*\s,-16*\s-0.02) {$x_t = [q; c_t]$};
    \node[grp, anchor=north] at (12*\s,-16*\s-0.02) {$x = [q; c]$};
    \node[grp, anchor=east] at (-0.02,-4*\s) {$x_t$};
    \node[grp, anchor=east] at (-0.02,-12*\s) {$x$};
  \end{scope}

  \begin{scope}[shift={(7.35,-0.25)}]
    \path[qcell=figInk] (0,0) rectangle ++(\s,\s);
    \node[grp, anchor=west] at (\s+0.05,0.5*\s) {query key (tint; never masked)};
    \path[dcell=figInk] (0,-0.3) rectangle ++(\s,\s);
    \node[grp, anchor=west] at (\s+0.05,-0.3+0.5*\s) {docid key (paradigm colour)};
    \path[off] (0,-0.6) rectangle ++(\s,\s);
    \node[grp, anchor=west] at (\s+0.05,-0.6+0.5*\s) {no attention};
  \end{scope}
\end{tikzpicture}
  \caption{Every paradigm sees the whole query; they differ in what each
    docid position sees of the docid. Attention masks for a 4-token query
    $q$ and a 4-level docid $c$ (row attends to column). (a) AR: the
    decoder cross-attends to the encoded query and is causal over $c$.
    (b) MDLM: query and partly masked docid $z_t$ form one sequence with
    full attention. (c) Block diffusion (block size 2): the doubled stream
    of BD3-LM~\cite{arriola2025bd3lm} with $x = [q; c]$; each noisy docid
    block attends to itself, the clean query and the clean earlier blocks.}
  \label{fig:conditioning}
\end{figure}
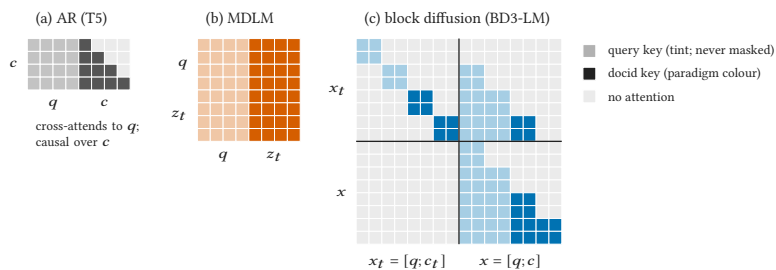
\clearpage

\section{What the Models Learn}
\begin{table}[!htbp]
  \centering
  \caption{Per-token accuracy (\%) where AR and diffusion see the same
    context (\NQ{} eval queries): level 1 from the query alone,
    level 16 with the other 15 levels given. Diffusion: range over MDLM,
    BD8, BD4. On training queries AR is near 100 at every level; diffusion
    PQ rises from 88--90 (level 1) to 90--97 (level 16), RQ falls from
    88--89 to 79--85.}
  \label{tab:matched}
  \tablesep
  \tablefont
  \begin{tabular}{@{}l
      S[table-format=2.0] S[table-format=2.0]
      S[table-format=2.0] S[table-format=2.0]@{}}
    \toprule
    & \multicolumn{2}{c}{Level 1} & \multicolumn{2}{c}{Level 16} \\
    \cmidrule(lr){2-3}\cmidrule(l){4-5}
    & {\PQ{}} & {\RQ{}} & {\PQ{}} & {\RQ{}} \\
    \midrule
    \pdot{arGray}\AR{} & \best{61} & \best{70} & \best{79} & \best{82} \\
    Diffusion & {49--50} & {59--60} & {67--78} & {54--58} \\
    \bottomrule
  \end{tabular}
\end{table}

\begin{table}[!htbp]
  \centering
  \caption{\NQ{} Hit@$k$ (\%) split by whether the gold document is the gold
    of a training query (seen: 6,075 queries; unseen: 1,755 of 7,830). Each
    paradigm at its best decoding (\cref{tab:snap}): AR trie beam, MDLM
    one-pass scoring, BD8 and BD4 chain-rule rescoring. Bold: best per column
    in the top block.
    $^g$10 generated queries per training document added at the same 16M
    examples, so real queries are seen $\sim$4.5$\times$ less often; paired
    change against MDLM above, seen / unseen: $-13.5$ [$-14.8$, $-12.1$] /
    $+1.1$ [$+0.2$, $+2.1$].}
  \label{tab:seen}
  \tablesep
  \tablefont
  \begin{tabular}{@{}l
      S[table-format=2.1] S[table-format=1.1] S[table-format=2.1] S[table-format=2.1]
      S[table-format=2.1] S[table-format=1.1] S[table-format=2.1] S[table-format=2.1]@{}}
    \toprule
    & \multicolumn{4}{c}{\PQ{}} & \multicolumn{4}{c}{\RQ{}} \\
    \cmidrule(lr){2-5}\cmidrule(l){6-9}
    & {seen} & \multicolumn{3}{c}{unseen} & {seen} & \multicolumn{3}{c}{unseen} \\
    \cmidrule(lr){3-5}\cmidrule(l){7-9}
    Paradigm & {@1} & {@1} & {@10} & {@100} & {@1} & {@1} & {@10} & {@100} \\
    \midrule
    \pdot{arGray}\AR{} (beam)            & \best{62.0} & \best{3.2} & 8.7 & 20.6 & \best{57.9} & 2.1 & 7.3 & 18.5 \\
    \pdot{mdlmOrange}\MDLM{} (one-pass)  & 57.2 & 2.7 & \best{16.4} & \best{40.7} & 53.3 & \best{2.9} & \best{19.7} & \best{37.2} \\
    \pdot{bdeightBlue}\BDeight{} (chain) & 53.9 & 1.8 & 6.8 & 40.5 & 52.1 & 1.9 & 16.4 & 36.1 \\
    \pdot{bdfourGreen}\BDfour{} (chain)  & 52.0 & 1.5 & 5.7 & 40.1 & 50.2 & 1.9 & 14.7 & 34.0 \\
    \addlinespace
    \multicolumn{9}{@{}l}{\emph{Real + generated queries, same budget}$^g$} \\
    \pdot{mdlmOrange}\MDLM{} (one-pass)  & 43.7 & 3.8 & {--} & {--} & {--} & {--} & {--} & {--} \\
    \bottomrule
  \end{tabular}
\end{table}

\clearpage

\section{Generation and Decoding}
\begin{table}[!htbp]
\footnotesize
  \centering
  \caption{The decodings compared, all on the same trained models. AR's trie
    beam and diffusion's generate-and-match give \cref{tab:arms}.}
  \label{tab:decodes}
  \tablefont
  \begin{tabular*}{\linewidth}{@{\extracolsep{\fill}}lll@{}}
    \toprule
    Decoding & Generates & Docid chosen by \\
    \midrule
    \multicolumn{3}{@{}l}{\emph{AR}} \\
    greedy, no trie     & 1 code, free        & exact match to a docid \\
    greedy, trie        & 1 code, in the trie & the trie (always a docid) \\
    beam 100, trie      & 100 codes, in the trie & summed log-probability \\
    \addlinespace
    \multicolumn{3}{@{}l}{\emph{Diffusion}} \\
    one sample          & 1 stochastic sample  & code matching \\
    argmax              & 1 argmax code        & code matching \\
    generate-and-match  & 1 argmax code        & code matching, reranked by $s(d)$ \\
    best-of-8           & 8 stochastic samples & pooled code matching, by $s(d)$ \\
    one-pass scoring    & none                 & $s(d)$ over the whole corpus \\
    chain-rule rescoring & none                & one-pass top 20, by chain-rule score \\
    \bottomrule
  \end{tabular*}
\end{table}

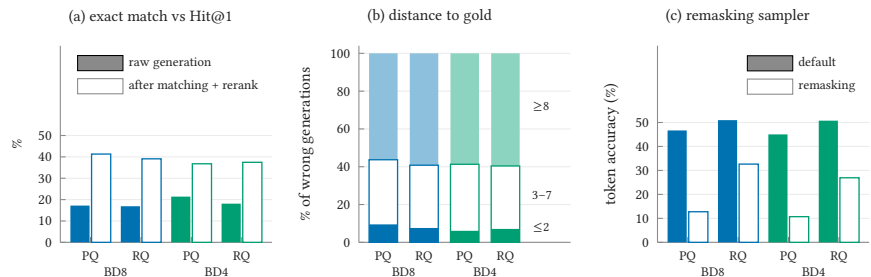
\begin{figure}[!htbp]
  \centering
  \pgfplotstableread[col sep=comma, row sep=\\]{x,cell,raw,reported,near,far,default,remask,revised\\
2,BD8-PQ,17.2,41.32,9.5,56.3,46.68,12.76,0.75\\
3,BD8-RQ,16.9,39.09,7.6,59.2,50.98,32.64,0.30\\
4,BD4-PQ,21.4,36.77,6.1,58.7,45.03,10.71,0.68\\
5,BD4-RQ,18.1,37.46,7.1,59.6,50.76,26.93,0.33\\
}\datageneration
\begin{tikzpicture}
  \pgfplotsset{
    rawbar/.style={ybar, bar width=0.38, bar shift=-0.21, draw=none, fill=#1},
    repbar/.style={ybar, bar width=0.38, bar shift=0.21, draw=#1,
      line width=0.5pt, fill=white},
    distbar/.style={ybar, bar width=0.7, bar shift=0, draw=none, forget plot},
    cellx/.style={xmin=1.4, xmax=5.6, xtick={2,3,4,5},
      xticklabels={PQ,RQ,PQ,RQ}, clip=false},
  }
  \begin{groupplot}[
    group style={group size=3 by 1, horizontal sep=1.15cm},
    figaxis, height=2.6cm,
  ]
    \nextgroupplot[title={(a) exact match vs Hit@1}, width=2.8cm, cellx,
      ymin=0, ymax=92, ytick={0,10,20,30,40,50}, ylabel={\%},
      legend style={at={(1,1)}, anchor=north east}]
      \addlegendimage{area legend, draw=none, fill=figMuted}
      \addlegendentry{raw generation}
      \addlegendimage{area legend, draw=figMuted, fill=white}
      \addlegendentry{after matching + rerank}
      \addplot[rawbar=bdeightBlue, forget plot, restrict x to domain=1.5:3.5] table[x=x, y=raw]{\datageneration};
      \addplot[rawbar=bdfourGreen, forget plot, restrict x to domain=3.5:5.5] table[x=x, y=raw]{\datageneration};
      \addplot[repbar=bdeightBlue, forget plot, restrict x to domain=1.5:3.5] table[x=x, y=reported]{\datageneration};
      \addplot[repbar=bdfourGreen, forget plot, restrict x to domain=3.5:5.5] table[x=x, y=reported]{\datageneration};
      \node[font=\tiny, text=figInk, yshift=-11pt] at (axis cs:2.5,0) {BD8};
      \node[font=\tiny, text=figInk, yshift=-11pt] at (axis cs:4.5,0) {BD4};
    \nextgroupplot[title={(b) distance to gold}, width=2.8cm, cellx,
      xmax=6.6, ymin=0, ymax=104, ytick={0,20,40,60,80,100},
      ylabel={\% of wrong generations}]
      \addplot[distbar, fill=bdeightBlue!45, restrict x to domain=1.5:3.5] table[x=x, y expr=100]{\datageneration};
      \addplot[distbar, draw=bdeightBlue, fill=white, line width=0.5pt, restrict x to domain=1.5:3.5] table[x=x, y expr=100-\thisrow{far}]{\datageneration};
      \addplot[distbar, fill=bdeightBlue, restrict x to domain=1.5:3.5] table[x=x, y=near]{\datageneration};
      \addplot[distbar, fill=bdfourGreen!45, restrict x to domain=3.5:5.5] table[x=x, y expr=100]{\datageneration};
      \addplot[distbar, draw=bdfourGreen, fill=white, line width=0.5pt, restrict x to domain=3.5:5.5] table[x=x, y expr=100-\thisrow{far}]{\datageneration};
      \addplot[distbar, fill=bdfourGreen, restrict x to domain=3.5:5.5] table[x=x, y=near]{\datageneration};
      \node[font=\tiny, text=figInk, anchor=south west] at (axis cs:5.45,1) {$\le$2};
      \node[font=\tiny, text=figInk, anchor=west] at (axis cs:5.45,25) {3--7};
      \node[font=\tiny, text=figInk, anchor=west] at (axis cs:5.45,72) {$\ge$8};
      \node[font=\tiny, text=figInk, yshift=-11pt] at (axis cs:2.5,0) {BD8};
      \node[font=\tiny, text=figInk, yshift=-11pt] at (axis cs:4.5,0) {BD4};
    \nextgroupplot[title={(c) remasking sampler}, width=2.8cm, cellx,
      ymin=0, ymax=82, ytick={0,10,20,30,40,50},
      ylabel={token accuracy (\%)},
      legend style={at={(1,1)}, anchor=north east}]
      \addlegendimage{area legend, draw=none, fill=figMuted}
      \addlegendentry{default}
      \addlegendimage{area legend, draw=figMuted, fill=white}
      \addlegendentry{remasking}
      \addplot[rawbar=bdeightBlue, forget plot, restrict x to domain=1.5:3.5] table[x=x, y=default]{\datageneration};
      \addplot[rawbar=bdfourGreen, forget plot, restrict x to domain=3.5:5.5] table[x=x, y=default]{\datageneration};
      \addplot[repbar=bdeightBlue, forget plot, restrict x to domain=1.5:3.5] table[x=x, y=remask]{\datageneration};
      \addplot[repbar=bdfourGreen, forget plot, restrict x to domain=3.5:5.5] table[x=x, y=remask]{\datageneration};
      \node[font=\tiny, text=figInk, yshift=-11pt] at (axis cs:2.5,0) {BD8};
      \node[font=\tiny, text=figInk, yshift=-11pt] at (axis cs:4.5,0) {BD4};
  \end{groupplot}
\end{tikzpicture}
  \caption{What block diffusion generates on \NQ{}. (a) Exact-match rate of
    the raw generation against Hit@1 after code matching and rerank.
    (b) Wrong raw generations by the number of wrong levels out of 16
    ($\le$2, 3--7, $\ge$8). (c) Per-token accuracy on 2,000 eval queries with
    the default sampler (filled) and a predict-then-noise sampler that
    overwrites every committed token at every step (open).}
  \label{fig:generation}
\end{figure}
\begin{table}[!htbp]
  \centering
  \caption{One-pass scoring and chain-rule rescoring, supporting measurements
    (\NQ{}). (a) Per-level accuracy with every level masked, and MDLM's
    chain-rule rescoring on a wider shortlist. (b) Other orders for
    chain-rule rescoring: BD scored block by block (earlier blocks revealed),
    MDLM with levels revealed right to left or most-confident first; top 20
    of one-pass scoring in every case. $\Delta$ Hit@1 in points, paired 95\%
    interval in brackets.}
  \label{tab:snapmore}
  \setlength{\tabcolsep}{2.5pt}
  \scriptsize
  \begin{adjustbox}{max width=\linewidth}
\begin{tabular}{@{}lcc@{}}
    \multicolumn{3}{@{}l}{(a) Supporting measurements} \\
    \toprule
    & \PQ{} & \RQ{} \\
    \midrule
    Per-level accuracy, all levels masked & \multicolumn{2}{c}{MDLM 47--52\% at every level; BD first block 49--50\%;} \\
    & \multicolumn{2}{c}{first level of each later block $\le$2\%, 0--41\% within it} \\
    MDLM chain rule on the top 100 (vs top 20): Hit@1 & 41.5 ($-0.4$ [$-0.6$, $-0.2$]) & 41.4 (flat) \\
    \quad top-1 agreement with the top 20 & 92\% & 96\% \\
    \bottomrule
  \end{tabular}
\end{adjustbox}

  \medskip
  \begin{adjustbox}{max width=\linewidth}
\begin{tabular}{@{}lllccc@{}}
    \multicolumn{6}{@{}l}{(b) Other chain-rule orders} \\
    \toprule
    & & Order & Hit@1 & vs one-pass & vs chain rule \\
    \midrule
    \BDfour{}  & \PQ{} & block by block & 41.1 & $+3.9$ [3.1, 4.7] & $+0.3$ [$-0.1$, 0.7] \\
               & \RQ{} & block by block & 39.4 & $+1.8$ [1.1, 2.5] & $+0.0$ [$-0.2$, 0.3] \\
    \BDeight{} & \PQ{} & block by block & 43.1 & $+1.6$ [1.0, 2.3] & $+0.9$ [0.5, 1.4] \\
               & \RQ{} & block by block & 40.9 & $+0.6$ [0.0, 1.2] & $+0.0$ [$-0.3$, 0.3] \\
    \MDLM{}    & \PQ{} & right to left  & 41.9 & $-3.1$ [$-3.6$, $-2.5$] & $+0.0$ [$-0.5$, 0.6] \\
               &       & confidence first & 41.3 & $-3.7$ [$-4.3$, $-3.0$] & $-0.6$ [$-1.1$, $-0.2$] \\
               & \RQ{} & right to left  & 41.1 & $-0.9$ [$-1.3$, $-0.5$] & $-0.4$ [$-0.8$, 0.1] \\
               &       & confidence first & 41.2 & $-0.8$ [$-1.2$, $-0.3$] & $-0.2$ [$-0.6$, 0.1] \\
    \bottomrule
  \end{tabular}
\end{adjustbox}
\end{table}

\clearpage

\section{Error Taxonomy}
\begin{table}[!htbp]
  \colortablestrue
  \renewcommand{\heat}[1]{\cellcolor{okPurple!\fpeval{round(0.6*#1)}}#1}
  \centering
  \caption{Error taxonomy (\%). Outcome: share of queries by the rank of the
    gold document. Cause: share of rank-1 misses by what the wrong top-1
    document is (first match wins, \cref{sec:setup}): near-duplicate of the
    gold, at most 2 of 16 levels wrong, topical, or unrelated. Shading grows
    with the share.}
  \label{tab:errors}
  \setlength{\tabcolsep}{2.5pt}
  \tablefont
  \begin{tabular}{@{}llcccc@{\hskip 6pt}cccc@{}}
    \toprule
    & & \multicolumn{4}{c}{Outcome (rank of gold)} & \multicolumn{4}{c}{Cause of a miss} \\
    \cmidrule(lr){3-6}\cmidrule(l){7-10}
    Paradigm & Decode & 1 & 2--10 & 11--100 & $>$100 & dup. & near & topic. & unrel. \\
    \midrule
    \multicolumn{10}{@{}l}{\emph{\NQ{}, \PQ{}}} \\
    \pdot{arGray}\textcolor{arGray}{\AR{}} & beam & \heat{48.8} & \heat{15.0} & \heat{8.7} & \heat{27.5} & \heat{0.8} & \heat{1.6} & \heat{52.2} & \heat{45.4} \\
    \pdot{mdlmOrange}\textcolor{mdlmOrange}{\MDLM{}} & one-pass & \heat{45.0} & \heat{18.5} & \heat{14.3} & \heat{22.3} & \heat{0.7} & \heat{3.1} & \heat{55.6} & \heat{40.6} \\
    \pdot{bdeightBlue}\textcolor{bdeightBlue}{\BDeight{}} & gen.\,\&\,match & \heat{41.3} & \heat{17.7} & \heat{7.2} & \heat{33.9} & \heat{0.5} & \heat{3.2} & \heat{54.2} & \heat{42.0} \\
    \pdot{bdeightBlue}\textcolor{bdeightBlue}{\BDeight{}} & chain rule & \heat{42.2} & \heat{18.4} & \heat{16.8} & \heat{22.5} & \heat{0.8} & \heat{2.8} & \heat{55.2} & \heat{41.2} \\
    \pdot{bdfourGreen}\textcolor{bdfourGreen}{\BDfour{}} & gen.\,\&\,match & \heat{36.8} & \heat{18.2} & \heat{7.0} & \heat{38.0} & \heat{0.5} & \heat{2.9} & \heat{52.2} & \heat{44.4} \\
    \pdot{bdfourGreen}\textcolor{bdfourGreen}{\BDfour{}} & chain rule & \heat{40.7} & \heat{18.3} & \heat{17.4} & \heat{23.6} & \heat{0.7} & \heat{2.7} & \heat{54.2} & \heat{42.4} \\
    \multicolumn{10}{@{}l}{\emph{\NQ{}, \RQ{}}} \\
    \pdot{arGray}\textcolor{arGray}{\AR{}} & beam & \heat{45.4} & \heat{13.5} & \heat{8.8} & \heat{32.4} & \heat{1.1} & \heat{0.1} & \heat{43.9} & \heat{54.8} \\
    \pdot{mdlmOrange}\textcolor{mdlmOrange}{\MDLM{}} & one-pass & \heat{42.0} & \heat{18.1} & \heat{10.6} & \heat{29.3} & \heat{0.9} & \heat{0.1} & \heat{50.1} & \heat{48.9} \\
    \pdot{bdeightBlue}\textcolor{bdeightBlue}{\BDeight{}} & gen.\,\&\,match & \heat{39.1} & \heat{10.8} & \heat{0.8} & \heat{49.3} & \heat{0.7} & \heat{0.1} & \heat{46.8} & \heat{52.4} \\
    \pdot{bdeightBlue}\textcolor{bdeightBlue}{\BDeight{}} & chain rule & \heat{40.9} & \heat{18.3} & \heat{10.9} & \heat{29.8} & \heat{0.7} & \heat{0.1} & \heat{50.8} & \heat{48.4} \\
    \pdot{bdfourGreen}\textcolor{bdfourGreen}{\BDfour{}} & gen.\,\&\,match & \heat{37.5} & \heat{10.9} & \heat{0.9} & \heat{50.7} & \heat{0.7} & \heat{0.1} & \heat{46.1} & \heat{53.1} \\
    \pdot{bdfourGreen}\textcolor{bdfourGreen}{\BDfour{}} & chain rule & \heat{39.3} & \heat{17.3} & \heat{11.5} & \heat{31.9} & \heat{0.8} & \heat{0.1} & \heat{51.0} & \heat{48.1} \\
    \multicolumn{10}{@{}l}{\emph{\NQ{}, random codes}} \\
    \pdot{arGray}\textcolor{arGray}{\AR{}} & beam & \heat{40.0} & \heat{11.3} & \heat{6.0} & \heat{42.6} & \heat{1.3} & \heat{0.0} & \heat{38.2} & \heat{60.5} \\
    \pdot{bdeightBlue}\textcolor{bdeightBlue}{\BDeight{}} & gen.\,\&\,match & \heat{35.2} & \heat{4.1} & \heat{0.1} & \heat{60.6} & \heat{1.0} & \heat{0.0} & \heat{35.6} & \heat{63.4} \\
    \pdot{bdfourGreen}\textcolor{bdfourGreen}{\BDfour{}} & gen.\,\&\,match & \heat{32.8} & \heat{2.8} & \heat{0.1} & \heat{64.2} & \heat{0.8} & \heat{0.0} & \heat{27.8} & \heat{71.4} \\
    \midrule
    \multicolumn{10}{@{}l}{\emph{\MSM{}, \PQ{}}} \\
    \pdot{arGray}\textcolor{arGray}{\AR{}} & beam & \heat{24.9} & \heat{30.2} & \heat{25.7} & \heat{19.2} & \heat{3.0} & \heat{4.4} & \heat{73.6} & \heat{18.9} \\
    \pdot{mdlmOrange}\textcolor{mdlmOrange}{\MDLM{}} & one-pass & \heat{16.3} & \heat{31.1} & \heat{28.3} & \heat{24.3} & \heat{2.4} & \heat{5.8} & \heat{75.1} & \heat{16.7} \\
    \pdot{bdeightBlue}\textcolor{bdeightBlue}{\BDeight{}} & gen.\,\&\,match & \heat{17.3} & \heat{25.4} & \heat{18.4} & \heat{38.9} & \heat{2.1} & \heat{5.7} & \heat{75.9} & \heat{16.3} \\
    \pdot{bdeightBlue}\textcolor{bdeightBlue}{\BDeight{}} & chain rule & \heat{12.9} & \heat{32.3} & \heat{33.3} & \heat{21.5} & \heat{2.1} & \heat{7.2} & \heat{71.9} & \heat{18.8} \\
    \pdot{bdfourGreen}\textcolor{bdfourGreen}{\BDfour{}} & gen.\,\&\,match & \heat{14.0} & \heat{27.0} & \heat{16.3} & \heat{42.7} & \heat{1.9} & \heat{5.0} & \heat{74.7} & \heat{18.4} \\
    \pdot{bdfourGreen}\textcolor{bdfourGreen}{\BDfour{}} & chain rule & \heat{14.4} & \heat{33.8} & \heat{31.2} & \heat{20.7} & \heat{2.3} & \heat{5.9} & \heat{73.0} & \heat{18.8} \\
    \multicolumn{10}{@{}l}{\emph{\MSM{}, \RQ{}}} \\
    \pdot{arGray}\textcolor{arGray}{\AR{}} & beam & \heat{21.3} & \heat{27.4} & \heat{22.2} & \heat{29.2} & \heat{1.7} & \heat{0.0} & \heat{70.9} & \heat{27.4} \\
    \pdot{mdlmOrange}\textcolor{mdlmOrange}{\MDLM{}} & one-pass & \heat{18.8} & \heat{32.1} & \heat{24.0} & \heat{25.1} & \heat{2.3} & \heat{0.2} & \heat{78.5} & \heat{19.1} \\
    \pdot{bdeightBlue}\textcolor{bdeightBlue}{\BDeight{}} & gen.\,\&\,match & \heat{16.3} & \heat{23.3} & \heat{3.3} & \heat{57.1} & \heat{2.1} & \heat{0.3} & \heat{74.9} & \heat{22.8} \\
    \pdot{bdeightBlue}\textcolor{bdeightBlue}{\BDeight{}} & chain rule & \heat{19.4} & \heat{30.4} & \heat{23.9} & \heat{26.2} & \heat{2.3} & \heat{0.3} & \heat{77.9} & \heat{19.5} \\
    \pdot{bdfourGreen}\textcolor{bdfourGreen}{\BDfour{}} & gen.\,\&\,match & \heat{15.8} & \heat{21.3} & \heat{3.3} & \heat{59.5} & \heat{2.2} & \heat{0.1} & \heat{74.3} & \heat{23.4} \\
    \pdot{bdfourGreen}\textcolor{bdfourGreen}{\BDfour{}} & chain rule & \heat{17.9} & \heat{32.5} & \heat{22.2} & \heat{27.4} & \heat{2.3} & \heat{0.3} & \heat{76.6} & \heat{20.8} \\
    \multicolumn{10}{@{}l}{\emph{\MSM{}, random codes}} \\
    \pdot{arGray}\textcolor{arGray}{\AR{}} & beam & \heat{0.2} & \heat{0.4} & \heat{1.0} & \heat{98.4} & \heat{0.2} & \heat{0.0} & \heat{5.7} & \heat{94.0} \\
    \pdot{bdeightBlue}\textcolor{bdeightBlue}{\BDeight{}} & gen.\,\&\,match & \heat{6.2} & \heat{1.0} & \heat{0.1} & \heat{92.7} & \heat{1.6} & \heat{0.0} & \heat{20.1} & \heat{78.4} \\
    \pdot{bdfourGreen}\textcolor{bdfourGreen}{\BDfour{}} & gen.\,\&\,match & \heat{5.7} & \heat{0.7} & \heat{0.0} & \heat{93.6} & \heat{0.7} & \heat{0.0} & \heat{14.4} & \heat{84.9} \\
    \bottomrule
  \end{tabular}

\end{table}

\begin{table}[!htbp]
  \centering
  \caption{The largest groups of documents sharing one code. Id.:
    identifier; size: documents in the group; share with an identical title;
    share with an empty body; mean pairwise TF-IDF cosine (random pairs in
    parentheses); topic.}
  \label{tab:codegroups}
  \tablesep
  \tablefont
  \begin{tabular}{@{}lcccccc@{}}
    \toprule
    & Id. & Size & \shortstack{Same\\title (\%)} & \shortstack{Empty\\text (\%)} & \shortstack{Cosine\\(random)} & Topic \\
    \midrule
    \NQ{} & \PQ{} & 75 & 1 & 0 & .745 (.041) & visa-policy pages \\
    \NQ{} & \PQ{} & 70 & 1 & 0 & .226 (.041) & FIFA World Cup \\
    \NQ{} & \PQ{} & 59 & 2 & 0 & .224 (.041) & shared Hinduism nav box \\
    \MSM{} & \RQ{} & 512 & 92 & 100 & .000 (.015) & empty pages \\
    \MSM{} & \PQ{} & 737 & 92 & 100 & .000 (.015) & empty pages \\
    \bottomrule
  \end{tabular}
\end{table}

\clearpage

\end{document}